\documentclass[reprint,aps,pra,showpacs,floatfix]{revtex4-1}

\usepackage{amsmath}
\usepackage{graphicx}
\usepackage{hyperref}
\usepackage{dcolumn}
\usepackage{color}
\hypersetup{colorlinks=true,citecolor=blue,urlcolor=blue,linkcolor=blue}

\newcommand{\Z}{Z_{\rm eff}}
\newcommand{\Zr}{Z_{\rm eff}^{({\rm res})}}
\newcommand{\eps}{\varepsilon}

\begin{document}

\title{Mode coupling and multiquantum vibrational excitations in Feshbach-resonant positron annihilation in molecules}

\author{G. F. Gribakin}\email{g.gribakin@qub.ac.uk}
\affiliation{School of Mathematics and Physics, Queen's University Belfast,
Belfast BT7 1NN, United Kingdom}
\author{J. F. Stanton}\email{jfstanton137@gmail.com}
\affiliation{Quantum Theory Project, Departments of Chemistry and Physics, University of Florida, Gainesville, FL 32611, USA}
\author{J. R. Danielson}
\author{M. R. Natisin}
\author{C. M. Surko}\email{csurko@ucsd.edu}
\affiliation{Department of Physics, University of California, San Diego, La Jolla, California 92093, USA}
\date{\today}

\begin{abstract}
The dominant mechanism of low-energy positron annihilation in polyatomic molecules is through positron capture in vibrational Feshbach resonances (VFR). In this paper we investigate theoretically the effect of anharmonic terms in the vibrational Hamiltonian on the positron annihilation rates. Such interactions enable positron capture in VFRs associated with multiquantum vibrational excitations, leading to enhanced annihilation. Mode coupling can also lead to faster depopulation of VFRs, thereby reducing their contribution to the annihlation rates. To analyze this complex picture, we use coupled-cluster methods to calculate the anharmonic vibrational spectra and dipole transition amplitudes for chloroform, chloroform-$d_1$, 1,1-dichloroethylene, and methanol, and use these data to compute positron resonant annihilation rates for these molecules. Theoretical predictions are compared with the annihilation rates measured as a function of incident positron energy. The results demonstrate the importance of mode coupling in both enhancement and suppression of the VFR. There is also experimental evidence for the direct excitation of multimode VFR.
Their contribution is analyzed using a statistical approach, with an outlook towards more accurate treatment of this phenomenon.
\end{abstract}

\pacs{34.80.Uv,34.80.Lx,33.20.Tp,78.70.Bj}

\maketitle

\section{Introduction}

It has been firmly established over the past fifteen years that positron annihilation with most polyatomic molecules proceeds through formation of vibrational Feshbach resonances (VFR) \cite{RMP2010}. Resonant annihilation strongly enhances positron-molecule annihilation rates, compared to those of direct, ``in-flight'' annihilation
\cite{Gribakin2000,Gribakin2001}. It also results in a characteristic dependence of the annihilation rates on the positron energy, which could be measured using a trap-based positron beam \cite{Gilbert2002}. These annihilation spectra carry signatures of the vibrational level structure of the molecule, allowing measurements of positron-molecule binding energies \cite{Barnes2003,Danielson2009,Danielson2010,Danielson2012a,Danielson2012b}.

For small polyatomic molecules with infrared-active vibrational modes,  such as methyl halides, there is a theory that enables calculations of resonant annihilation \cite{Gribakin2006}. According to this theory, the probability of positron capture into individual VFRs is determined by the corresponding vibrational transition dipole amplitudes. The theory contains one free parameter, namely, the positron-molecule binding energy $\eps _b$, a quantity that has proved to be difficult to predict theoretically. Realistic values of the binding energy have been obtained for strongly polar molecules (e.g., by using a configuration interaction scheme with singly and doubly excited levels \cite{TKB11,TKB12,T14}). Yet, even in the best case (acetonitrile), the calculated and measured values of $\eps _b$ differ by 25\%.

However, the theory of Ref.~\cite{Gribakin2006} cannot describe strongly enhanced resonant peaks observed for larger polyatomic molecules \cite{YS08,GL09}. Here the mode-based resonances act as doorways \cite{GG04} into multimode vibrational states, which leads to longer positron capture times due to intramolecular vibrational energy redistribution (IVR) \cite{RMP2010}. The theory also has difficulty in describing the annihilation rates in smaller molecules, such as ethylene, in which dipole-forbidden vibrational excitations or mode-mixing effects appear to be important \cite{YGLS08,YS08a,Gribakin2010,Natisin2017}. One of the manifestations of mode mixing is reduction of the magnitudes of some VFRs due to vibrationally inelastic escape of the positron (i.e., detachment of the positron by de-excitation of a mode other than that involved in its capture). So far, such effects have only been included phenomenologically \cite{GL09,Jones2013,Danielson2013}, by multiplying the contributions of mode-based VFR by scaling factors that can be greater or smaller than unity.

This paper is the first attempt to account for the contributions of multiquantum vibrational resonances and mode-mixing effects using a consistent theoretical framework. It is based on \textit{ab initio} calculations of the molecular vibrational eigenstates (inlcuding up to three-quantum excitations), taking account of anharmonic terms in the vibrational Hamiltonian. The coupling of the vibrational motion to the positron is described in the dipole approximation (as in Ref.~\cite{Gribakin2006}). This is appropriate for low positron energies and for molecules in which all modes are infrared-active. We provide a detailed analysis of the contributions of vibrational overtones and combinations to the resonant positron annihilation in chloroform and chloroform-$d_1$, 1,1-dichloroethylene and methanol. Comparison with experimental annihilation-rate data shows definitively that these states result in distinct features at specific positron energies. 

This is the first step towards a complete theory of positron annihilation in polyatomic molecules in which excitation of multimode vibrational states through the process of intramolecular vibrational energy redistribution (IVR) leads to dramatic enhancement of the annihilation rate.

\section{Theory}\label{sec:theory}

\subsection{Vibrational Feshbach resonances}

The positron-molecule VFR is a state in which the positron is bound to a
vibrationally excited molecule. The energy of this state relative to the molecular
ground state is given by
\begin{equation}\label{eq:eps_nu}
\eps _\nu =E_\nu -E_0-\eps_b,
\end{equation}
where $E_\nu $ and $E_0$ are the energies of the excited vibrational state
$\nu $ and ground state $0$ of the molecule, and $\eps_b$ is the positron-molecule
binding energy. The assumption in Eq.~(\ref{eq:eps_nu}) is that the positron binding does not affect the vibrational energy levels of the molecule, nor is the positron binding energy vibrational-state-dependent. The validity of these assumptions is borne by both extensive experimental data \cite{RMP2010} and calculations (see, e.g., Ref.~\cite{Kita2014}).

For $E_\nu -E_0>\eps _b$, the energy $\eps_\nu $ in Eq.~(\ref{eq:eps_nu}) is
positive, which means that this state is embedded in the positron-molecule
continuum, hence it is a quasibound state, or resonance. This resonance can be
populated when the energy $\eps $ of the positron
incident on the molecule (assumed to be in the ground state) is close to
$\eps _\nu $. The probability $P_\nu$ of populating state $\nu$ is described by the Breit-Wigner resonance profile \cite{LandauQM},
\begin{equation}\label{eq:BW}
P_\nu \propto \frac{\Gamma _\nu ^e}{(\eps -\eps _\nu )^2+\Gamma _\nu ^2/4},
\end{equation}
where $\Gamma _\nu ^e$ is the so-called elastic width (i.e., that corresponding to the entrance channel), and $\Gamma _\nu $ is the total width of the VFR (see Sec.~\ref{subsec:widths}). The total positron resonant annihilation cross section is given by the sum over the resonances \cite{LandauQM},
\begin{equation}\label{eq:sigma_a}
\sigma _a=\frac{\pi }{k^2}\sum _\nu \frac{\Gamma ^a\Gamma _\nu ^e}{(\eps -\eps _\nu )^2+\Gamma _\nu ^2/4}
\end{equation}
where $k$ is the incident positron momentum ($\eps =k^2/2$), and $\Gamma ^a$ is the annihilation width of the resonance, which is equal to the annihilation rate of the positron in the bound state (see Sec.~\ref{subsec:rate}). Here and elsewhere atomic units are used, in which $\hbar =e=m=1$, where $e$ is the elementary charge and $m$ is the electron or positron mass.

The magnitudes of the elastic widths are determined by the strength of coupling between the motion of the positron and the vibrational motion of the heavy nuclear framework of the molecule. Owing to the large mass difference, this coupling is small and the elastic widths usually do not exceed 0.1~meV \cite{Gribakin2006}. The total widths are also small, much smaller than the typical energy spread of the positron beam used for measuring the energy-dependent annihilation rates ($\Delta \eps \sim 40$~meV for a room-temperature buffer-gas trap-based beam \cite{Gilbert1997,Gilbert2002}, or 7~meV, for a cryogenic trap-based beam \cite{Natisin2016}). This means that the Breit-Wigner form (\ref{eq:BW}) can be replaced by the delta function of equivalent spectral weight,
\begin{equation}\label{eq:BW_delta}
2\pi \frac{\Gamma _\nu ^e}{\Gamma _\nu }\delta (\eps -\eps _\nu ).
\end{equation}
This shows that the contribution of VFR $\nu $ to the annihilation signal is proportional to the ratio $\Gamma _\nu ^e/\Gamma _\nu $. Here the factor $\Gamma _\nu^e$ determines the probability of positron capture in resonance $\nu $, while the factor $1/\Gamma _\nu$ is the positron lifetime in the resonant state, which determines the probability of its annihilation with one of the molecular electrons.
In large polyatomic molecules which have high vibrational level densities, VFRs can lead to orders-of-magnitude enhancements of the annihilation rates \cite{RMP2010}.

\subsection{Resonance widths}\label{subsec:widths}

The positron-molecule VFR described above can decay either by positron annihilation, or by positron re-emission back into the continuum. The latter process is driven by a downward vibrational transition which must supply the energy greater than $\eps _b$ in order to detach the positron. It is similar to the positron capture process in which the energy lost by the positron ($\eps + \eps _b$) is absorbed by the molecular vibrations. In this section we determine the rate, or partial width, corresponding to the positron detachment process. 

In the harmonic approximation, and assuming dipole coupling, the only vibrational states that are coupled are pairs which differ by one quantum in some mode. For positron capture by a ground-state molecule, this means that only the VFR corresponding to one-quantum excitations (the fundamental transitions) of infrared-active modes are allowed, as described in Ref.~\cite{Gribakin2006}.

When anharmonic terms are included in the molecular (Watson) Hamiltonian for a nonlinear polyatomic molecule, the wave function of the vibrational eigenstate $\nu $ can be written as
\begin{equation}\label{eq:Psi_nu}
\Psi _\nu (q)=\sum _nC^\nu_n\Phi _n(q).
\end{equation}
Here $\Phi _n$ are the harmonic basis states with vibrational
quantum numbers $n\equiv (n_1,\dots ,n_s)$ of the $s=3N-6$ normal modes,
$N$ being the number of atoms in the molecule, and
$n_i=0,\,1,\dots $ is the vibrational quantum number of mode $i$. The wave function in Eq.~(\ref{eq:Psi_nu}) depends on $s$ normal coordinates $q\equiv (q_1,\dots ,q_s)$. The (real) expansion coefficients $C^\nu_n$ and eigenstate energies $E_\nu $ can be obtained by diagonalization of the vibrational Hamiltonian matrix $H_{nn'}=\langle \Phi _{n'}|\hat H_{\rm vib}|\Phi _n\rangle $ (see Sec.~\ref{subsec:vib}).

To determine the partial widths of a VFR $\nu $, consider the process of positron emission from the corresponding bound state due to vibrational de-excitation of the molecule into some lower-lying final vibrational state $\nu '$. Working along the lines of Ref.~\cite{Gribakin2006}, we write the amplitude of this process as
\begin{align}
A_{\nu '\nu }({\bf k})&=\int \Psi _{\nu '}^*(q)e^{-i{\bf k}\cdot {\bf r}}
\frac{{\bf D}\cdot {\bf r}}{r^3}\Psi _\nu (q)\varphi _0({\bf r})d{\bf r}dq \notag \\
&=\langle \Psi _{\nu '}|{\bf D}|\Psi _\nu \rangle \cdot \int
\frac{{\bf r}}{r^3}e^{-i{\bf k}\cdot {\bf r}}\varphi _0({\bf r})d{\bf r}.
\label{eq:Anunu1}
\end{align}
Here ${\bf D}$ is the electric dipole moment operator of the molecule, ${\bf k}$
is the momentum of the ejected positron, whose magnitude is determined
by energy conservation,
\begin{equation}\label{eq:E_cons}
E_\nu - E_{\nu '}=\eps _b+k^2/2,
\end{equation}
and
\begin{equation}\label{eq:phi0}
\varphi _0({\bf r})=\sqrt{\frac{\kappa }{2\pi }}\frac{e^{-\kappa r}}{r}
\end{equation}
is the wave function of the weakly bound positron state with binding energy
$\eps _b=\kappa ^2/2$. The only difference between Eq.~(\ref{eq:Anunu1}) and the amplitude in Eq.~(6) of Ref.~\cite{Gribakin2006} is that the latter considered only transitions between the molecular ground state and any of the infrared-active fundamentals. As a result, instead of $\langle \Psi _{\nu '}|{\bf D}|\Psi _\nu \rangle $, the amplitude only involved $\langle \Phi _0|{\bf D}|\Phi _n\rangle $, which was assumed to be nonzero only for single-quantum excitations of a particular fundamental $i$, i.e., for $n=(0,\dots ,\,1,\dots ,0)\equiv 0[i^+]$, the ground state $0$ with one vibrational quantum added in mode $i$.

For anharmonic vibrational states (\ref{eq:Psi_nu}), the dipole amplitude in Eq.~(\ref{eq:Anunu1}) is given by
\[
\langle \Psi _{\nu '}|{\bf D}|\Psi _\nu \rangle =
\sum _{n,n'} C^{\nu '}_{n'}C^\nu_n\langle \Phi _{n'}|{\bf D}|\Phi _n\rangle .
\]
In the lowest-order approximation, the dipole operator ${\bf D}$ is a linear function of the normal coordinates $q$. In this case the dipole matrix element $\langle \Phi _{n'}|{\bf D}|\Phi _n\rangle $ is nonzero only if $n'$ and $n$ differ by one quantum of excitation in one of the modes. Given that the molecule is de-excited, this means that for $n=n_1,\dots ,n_i,\dots , n_s$, we have
$n'=n_1,\dots ,n_i-1,\dots , n_s\equiv n[i^-]$, so that
\begin{equation}\label{eq:PsiDPsi}
\langle \Psi _{\nu '}|{\bf D}|\Psi _\nu \rangle = \sum _n \sum _i
C^{\nu '}_{n[i^-]}C^\nu_n\langle \Phi _{n[i^-]}|{\bf D}|\Phi _n\rangle .
\end{equation}
The matrix element between two harmonic basis states is expressed in terms
of the oscillator matrix element for normal mode $i$,
\begin{equation*}\label{eq:dip_mat}
\langle \Phi _{n[i^-]}|{\bf D}|\Phi _n\rangle =\langle n_i-1|{\bf d}|n_i\rangle
=\sqrt{n_i}\langle 0|{\bf d}|1\rangle \equiv \sqrt{n_i} {\bf d}_i,
\end{equation*}
where ${\bf d}_i$ is the dipole amplitude for the excitation of mode $i$
from the ground to the first excited state (which determines the infrared
absorption intensity of this fundamental). Therefore, the transition amplitude between the vibrational states $\nu $ and $\nu'$ is given by
\begin{equation}\label{eq:Dnu'nu}
\langle \Psi _{\nu '}|{\bf D}|\Psi _\nu \rangle = \sum _n \sum _i
C^{\nu '}_{n[i^-]}C^\nu_n \sqrt{n_i} {\bf d}_i.
\end{equation}

Including quadratic and higher-order terms in the dependence of ${\bf D}$ on $q$ will give rise to two-quantum and higher corrections in the amplitude, Eq.~(\ref{eq:Dnu'nu})
(see Sec.~\ref{subsec:vib}). These terms will allow for the corresponding dipole transitions between the vibrational states even if the latter are described in the harmonic approximation.

The contribution of the transition $\nu \rightarrow \nu'$ to the width is
given by
\begin{equation*}\label{eq:Gnunu}
\Gamma _{\nu \rightarrow \nu'}=2\pi \int |A_{\nu '\nu}({\bf k})|^2
\delta \left(E_{\nu '}+\frac{k^2}{2}-E_\nu +\eps _b\right)\frac{d^3k}{(2\pi )^3}.
\end{equation*}
Proceeding in the same way as in Ref.~\cite{Gribakin2006}, we obtain
\begin{equation}\label{eq:Gnunu1}
\Gamma _{\nu \rightarrow \nu'}=\frac{16\omega _{\nu \nu'}D_{\nu'\nu }^2}{27}
h(\xi _{\nu \nu'}),  
\end{equation}
where $\omega _{\nu \nu'}=E_\nu -E_{\nu'}$, and
\begin{equation}\label{eq:h}
h(\xi )=\xi ^{3/2}(1-\xi )^{-1/2}
\left[ _2F_1\left(\frac{1}{2},1;\frac{5}{2};-\frac{\xi}{1-\xi }\right)
\right]^2,
\end{equation}
is a dimensionless function evaluated at
$\xi =\xi _{\nu \nu '}\equiv 1-\eps _b/\omega _{\nu \nu '}$ \footnote{The hypergeometric function in Eq.~(\ref{eq:h}) is
$_2F_1\left(\frac{1}{2},1;\frac{5}{2};-z^2\right )
=\frac32 z^{-2}[(1+z^2)z^{-1}\arctan z -1]$.}.
The ejection of the bound positron is possible only if
$\omega _{\nu \nu'}$ is greater than the positron binding energy, which means that $0<\xi _{\nu \nu'} <1$.

The probability of positron capture in the VFR $\nu $ by the ground-state
molecule is proportional to the elastic width
\begin{equation}\label{eq:Gnu0}
\Gamma _\nu ^e\equiv \Gamma _{\nu \rightarrow 0}=
\frac{16\omega _{\nu 0}D_{0\nu }^2}{27}h(\xi _{\nu 0}).
\end{equation}
Since the ground state of the molecule is largely immune to anharmonic state mixing, i.e., $\Psi_0=\Phi _0$, the corresponding dipole
amplitude is well approximated by
\begin{equation}\label{eq:D0nu}
\langle \Psi _0|{\bf D}|\Psi _\nu \rangle = 
\sum _{n} C^\nu_n\langle \Phi _0|{\bf D}|\Phi _n\rangle 
=\sum _i C^\nu_{0[i+]}{\bf d}_i.
\end{equation}
As with Eq.~(\ref{eq:Dnu'nu}), quadratic and higher terms in the dipole operator ${\bf D}$ can produce two-quantum, three-quantum, etc., corrections to the amplitude (\ref{eq:D0nu}).

The total width of the VFR is 
\begin{equation}\label{eq:Gam_nu}
\Gamma _\nu =\Gamma _\nu ^v+\Gamma ^a,
\end{equation}
where $\Gamma _\nu ^v=\sum _{\nu '}\Gamma _{\nu\rightarrow \nu'}$ is the width due to positron escape by vibrational de-excitation, and the sum is over all final states allowed by energy conservation, i.e., such that $E_{\nu'}<E_\nu -\eps_b$. The annihilation width $\Gamma ^a$ is given by the positron annihilation rate in the bound state,
\begin{equation}\label{eq:Gam_a}
\Gamma ^a=\pi r_0^2c\rho _{ep},
\end{equation}
where $\rho _{ep}$ is the electron-positron contact density in the bound state. For weakly bound positron-atom or positron-molecule states, it is given by
\begin{equation}\label{eq:rho_ep}
\rho_{ep}= (F/2\pi )\kappa ,
\end{equation}
with $F\approx 0.66$~a.u. \cite{Gribakin2001}. This means that $\Gamma ^a$ is determined by the positron binding energy and that $\Gamma ^a\propto \sqrt{\eps _b}$ (see also Ref.~\cite{Mitroy2002}) .
 
\subsection{Positron annihilation rate}\label{subsec:rate}

The positron annihilation rate in a gas of number density $n_m$ is $\lambda =\sigma _a vn_m$, where $v$ is the positron velocity. Conventionally, it is parameterized in terms of the Dirac annihilation rate in an uncorrelated electron gas, as
\begin{equation}\label{eq:ann_rate}
\lambda = \pi r_0^2c \Z n_m,
\end{equation}
where $r_0$ is the classical electron radius, $c$ is the speed of light, and $\Z$ is the effective number of electrons that contribute to positron annihilation on a given molecule \cite{Fraser68}. This interpretation of $\Z$ holds to some extent for simple molecules, like H$_2$, N$_2$ or O$_2$, in which positrons annihilate in flight and $\Z$ is comparable to the number of target electrons $Z$.

It has been known since the early works by Deutsch \cite{Deutsch51} that for polyatomic molecules and thermalized room-temperature positrons, the observed values of $\Z$ are much greater than $Z$ (see Ref.~\cite{RMP2010} and references therein). Such $\Z$ are unrelated to the number of target electrons, and are almost entirely due to resonant annihilation \cite{Gribakin2000,Gribakin2001}. In this case, using Eqs.~(\ref{eq:sigma_a}) and (\ref{eq:ann_rate}), one obtains
\begin{equation}\label{eq:Zeff_res}
\Zr (\eps )\equiv \frac{\sigma _a v}{\pi r_0^2c}=\frac{\Gamma ^a}{r_0^2ck}\sum _\nu \frac{\Gamma _\nu ^e}{(\eps -\eps _\nu )^2+\Gamma _\nu ^2/4}.
\end{equation}
Apart from large magnitudes, $\Zr$ has a characteristic energy dependence which  is strongly related to the molecular vibrational spectrum. This is a key feature of resonant annihilation \cite{Gilbert2002}. Critically, it allows measurements of positron-molecule binding energies from the downshifts of the resonances relative to the corresponding vibrational mode energies \cite{Barnes2003} [cf. Eq.~(\ref{eq:eps_nu})].

For molecules in which the density of VFR is not too high (e.g., with five or six atoms), the ability to resolve individual VFR in the $\Zr(\eps )$ spectrum is limited 
only by the energy resolution of the positron beam. To describe measured $\Z$ we need to convolve $\Zr(\eps )$ from Eq.~(\ref{eq:Zeff_res}) with the positron energy distribution function. For trap-based positron beams this distribution is Maxwellian in the transverse direction and approximately Gaussian in the longitudinal ($z$) direction,
\begin{equation}\label{eq:f}
f(\eps _\perp ,\eps _z)=\frac{1}{k_BT_\perp \sqrt{2\pi \sigma^2}}
\exp \left[ -\frac{\eps _\perp }{k_BT_\perp}
-\frac{(\eps _z -\epsilon )^2}{2\sigma^2}\right].
\end{equation}
Here  $\eps _\perp $ and $\eps _z$ are the transverse and longitudinal (or parallel) positron energies ($\eps =\eps _\perp +\eps _z)$, $\sigma $ is the root-mean-squared width of the parallel energy distribution (corresponding to a FWHM $\delta _z=\sigma\sqrt{8\ln 2}$), $\epsilon $ is the mean parallel energy of the positron beam, $k_B$ is Boltzmann's constant, and $T_\perp $ is the transverse positron temperature.

The experimentally measured normalized resonant annihilation rate,
\begin{equation}\label{eq:conv}
\bar{Z}_{\rm eff}^{(\rm res)} (\epsilon )=\int \Zr (\eps )f(\eps _\perp ,\eps _z) d\eps _\perp d \eps _z ,
\end{equation}
is calculated using Eq.~(\ref{eq:BW_delta}), which gives \cite{Gribakin2006}
\begin{equation}\label{eq:Zeff}
\bar{Z}_{\rm eff}^{(\rm res)}(\epsilon )=
2\pi^2\rho_{ep}\sum_\nu \frac{\Gamma _\nu^e}{k_\nu \Gamma _\nu}
\Delta (\epsilon -\eps _\nu),
\end{equation}
where $k_\nu =\sqrt{2\eps _\nu}$ is the resonance momentum, and
\begin{align}\label{eq:Delta}
\Delta (E)&=\frac{1}{2k_BT_\perp }
\exp \left[ \frac{\sigma ^2}{2(k_BT_\perp )^2}\right]
\exp \left(\frac{E}{k_BT_\perp }\right)\notag \\
&\times \left\{ 1+\Phi \left[ -\frac{1}{\sqrt{2}}
\left(\frac{E}{\sigma }+\frac{\sigma}{k_BT_\perp }\right)
\right]\right\},
\end{align}
with $\Phi (x)$ the standard error function \footnote{Note that a factor $1/2$ is missing in the expression for $\Delta (E)$ in Ref.~\cite{Gribakin2006}, Eq.~(10), although is was included in the calculations.}.

The function $\Delta (\epsilon -\eps _\nu)$ describes the shape of
a narrow resonance as observed with a trap-based positron
beam (see Fig.~1 in Ref.~\cite{Gribakin2006}). This function is asymmetric, with a low-energy tail due to the positron transverse energy content which allows it to access a resonance for $\epsilon <\eps _\nu$. For this reason the maxima of the resonant peaks described by Eq.~(\ref{eq:Zeff}) are also downshifted from the positions resonance energies. For typical room-temperature trap beam parameters $\delta _z=k_BT_\perp =25$~meV, this shift is about 12~meV, while the total FWHM of the resonance profile is 40~meV. Conversely, $\Delta (-E)$ gives the energy distribution of the positron beam with respect to its mean parallel energy.

Note that if any of the vibrational states is degenerate, its contribution to the sum in Eq.~(\ref{eq:Zeff}) should be multiplied by the corresponding degeneracy factor $g_\nu $.

\subsection{Calculation of vibrational eigenstates}\label{subsec:vib}

In order to calculate theoretical resonant $\Z$ spectra from Eq.~(\ref{eq:Zeff}) and compare them with experimental data, we use the vibrational state energies and transition dipole amplitudes obtained from
application of second-order vibrational perturbation theory (VPT2) \cite{VPT2}, in conjunction with potential energy surface (through quartic terms) and dipole surface (through cubic) terms calculated with coupled-cluster theory, the latter using the method known as CCSD(T) \cite{CC}. In cases that are free of Fermi and Darling-Dennison resonances (which occur in the presence of near degeneracy of levels differing by an odd or even number of vibrational quanta, respectively), this method gives an excellent picture of fundamental and two-quantum vibrational levels, with some work showing that the good treatment can also extend to three-quantum excitations in favorable cases \cite{THREEA,THREEB}. However, VPT2 does not treat four- and higher-quantum transitions, but these are unlikely to play an important role in the processes under investigation. (The exception is the contribution of statistical multimode resonant annihilation, which is known to provide a smooth background to the $\Z$ signal \cite{Jones2012}; see Sec. \ref{sec:disc}.)
The atomic natural orbital basis set known as ANO1 \cite{ANOA,ANOB} was used in the coupled-cluster calculations, in conjunction with the frozen core approximation. This information is sufficient to calculate the energies and dipole matrix elements needed to simulate the VFR position annihilation spectrum.

In methanol, however, there is a significant Fermi resonance between $\nu _3$ and $2\nu _{10}$ due to a near degeneracy of the zeroth-order harmonic levels (3012.8 and 3012.2~cm$^{-1}$, respectively). For the purposes of this paper, this resonance was removed by making a small adjustment to the harmonic frequencies of modes 3 and 10
(i.e., effectively a ``deperturbation'' of $\nu _3$). While this is simply an expedient, it is expected to have minimal effect on the qualitative nature of the spectral profile, which is what is being compared in this work. A detailed treatment of Fermi and Darling-Dennison resonances on the positron annihilation spectra is beyond the scope of this work.

\section{Results}\label{sec:res}

In this section, we apply the theory outlined in Sec.~\ref{sec:theory} to four molecules: chloroform (CHCl$_3$), chloroform-$d_1$ (CDCl$_3$), 1,1-dichloroethylene (C$_2$H$_2$Cl$_2$), and methanol (CH$_3$OH), and compare $\Z$ calculated \textit{ab initio} (for a fixed binding energy) with the measured annihilation rates. Specifically, we aim to identify features due to anharmonic corrections and mode mixing, such as VFR due to two- and three-quantum vibrational excitations, and VFR suppression due to vibrationally inelastic escape.

The resonant annihilation mechanism is operational only for the molecules that support positron bound states. The key parameters that determine the existence and strength of positron binding to neutral atoms and molecules, are the ionization energy $I$, dipole moment $\mu $ and dipole polarizability $\alpha _d$ \cite{Dzuba1995,Mitroy1999,Dzuba2010,Danielson2009,Danielson2010,Gribakin2015}. Their values for three of the molecules studied here are listed in Table \ref{tab:param}. 
(Deuteration is expected to have only a small effect on these properties.) Also listed are the binding energies inferred from the positions of the VFR in the measured $\Z$ spectra (see, e.g., Ref.~\cite{Danielson2012a}). Their uncertainty is typically $\pm 1$~meV, though is a little higher for 1,1-dichloroethylene ($\pm 3$~meV) \cite{Natisin_thesis}.
These data support the general trend of stronger positron binding for systems with larger $\alpha _d$ and $\mu $, the polarizability usually having a greater effect.

\begin{table}[ht]
\caption{Molecular ionization energies, dipole moments, and dipole polarizabilities
from Ref.~\cite{CRCHandbook} and positron binding energies inferred from the resonant $\Z$ spectra.}
\label{tab:param}
\begin{ruledtabular}
\begin{tabular}{ldcccc}
 & \multicolumn{1}{c}{$I$} & $\mu $ & $\alpha _d$ &  $\eps _b$ \\
Molecule & \multicolumn{1}{c}{(eV)} & (D) & (a.u.) & (meV) \\
\hline
Chloroform  CHCl$_3$ & 11.37 & 1.04 & 56, 64 & 40 \\
1,1-dichloroethylene C$_2$H$_2$Cl$_2$ & 9.81 & 1.34 & 53 & 35 \\
Methanol CH$_3$OH & 10.85 & 1.70 & 22 & 6
\end{tabular}
\end{ruledtabular}
\end{table}

\subsection{Chloroform}

The chloroform molecule (CHCl$_3$) has $C_{3v}$ symmetry. It can be viewed as an analog of methyl halides (CH$_3$X, $\text{X}=\text{F},\,\text{Cl},\, \text{Br}$) for which the observed $\Z$ are described well by mode-based VFR theory \cite{Gribakin2006}. Its ionization energy, dipole moment and dipole polarizability are close to those of CH$_3$Br ($I=10.54$~eV, $\mu =1.82$~D, and $\alpha _d\approx 40$~a.u.), so it is not surprising that the two molecules have similar positron binding energies \cite{Gribakin2006}.

The vibrational mode spectrum of chloroform is shown in Table \ref{tab:chcl3}. It lists values of the mode frequencies and squared transition dipole amplitudes
calculated as described in Sec.~\ref{subsec:vib}, in the harmonic approximation and with anharmonic corrections. It also shows selected frequency values from the NIST data tables \cite{NIST} and squared transition dipole moments that are based on measured infrared intensities \cite{Bishop1982}.

\begin{table}[ht]
\caption{Vibrational mode energies and dipole transition strengths 
$g_\nu D_{0\nu }^2$ ($g_\nu$ being the mode degeneracy) for chloroform. The notation $a[b]$ means $a\times 10^b$.}
\label{tab:chcl3}
\begin{ruledtabular}
\begin{tabular}{ccrrrccc}
\multicolumn{2}{l}{Mode and} & \multicolumn{3}{c}{$\omega $ (cm$^{-1}$)} & \multicolumn{3}{c}{$g_\nu D_{0\nu }^2$ (a.u.)} \\
\cline{3-5} \cline{6-8} 
\multicolumn{2}{l}{symmetry} & Harm. & Anh. & Sel.\footnote{Selected values from NIST tables~\cite{NIST}.} & Harm. & Anh. &
Exp.\footnote{Values derived from integrated intensity data \cite{Bishop1982}.}\\
\hline
$\nu _1$ & $a_1$ & 3180 & 3039 & 3034 & 3.69[-6] & 3.14[-7] & 7.04[-6]\\
$\nu _2$ & $a_1$ &  678 &  669 &  680 & 4.09[-4] & 4.41[-4] & 4.63[-4]\\
$\nu _3$ & $a_1$ &  369 &  364 &  363 & 4.97[-5] & 4.79[-5] & 4.94[-5]\\
$\nu _4$ & $e$   & 1245 & 1219 & 1220 & 2.06[-3] & 1.92[-3] & 2.41[-3]\\
$\nu _5$ & $e$   &  791 &  773 &  774 & 2.03[-2] & 2.03[-2] & 2.15[-2]\\
$\nu _6$ & $e$   &  262 &  259 &  261 & 3.30[-5] & 2.81[-5] & 3.09[-5]\\
\end{tabular}
\end{ruledtabular}
\end{table}

Including anharmonic effects, which are particularly large for the CH-stretch mode ($\nu _1$), brings all vibrational frequencies into close agreement with experiment. There is also a generally good accord between the calculated and measured values of the transition dipole moments. The largest uncertainty here is for the $\nu _1$ mode. It has a very small infrared absorption strength, and shows great sensitivity to the anharmonic corrections.

The calculated vibrational spectrum of the molecule includes one-, two- and three-quantum excitations from the ground state (a total of 220 states). When anharmonic effects are included, all these states can be coupled by dipole transition amplitudes. In practice, the coupling between states that differ by several vibrational quanta is small, although there are exceptions (see below).

Figure \ref{fig:chcl3} shows the calculated resonant $\Z$ averaged over the positron-beam energy distribution, as per Eq.~(\ref{eq:Zeff}), using different approximations. They are compared with the experimental data for chloroform \cite{Jones2013,Natisin_thesis}, which correct the earlier measurements \cite{Jones2012}. The positron binding energy for chloroform is taken to be $\eps _b=40$~meV \cite{Jones2013}, as this gives the best agreement for the resonance positions. The corresponding annihilation width from Eq.~(\ref{eq:Gam_a}) is $\Gamma ^a=6.95\times 10^{-9}$~a.u., which corresponds to the annihilation lifetime of 3.5~ns
for the bound positron. 

\begin{figure}
\includegraphics*[width=0.48\textwidth]{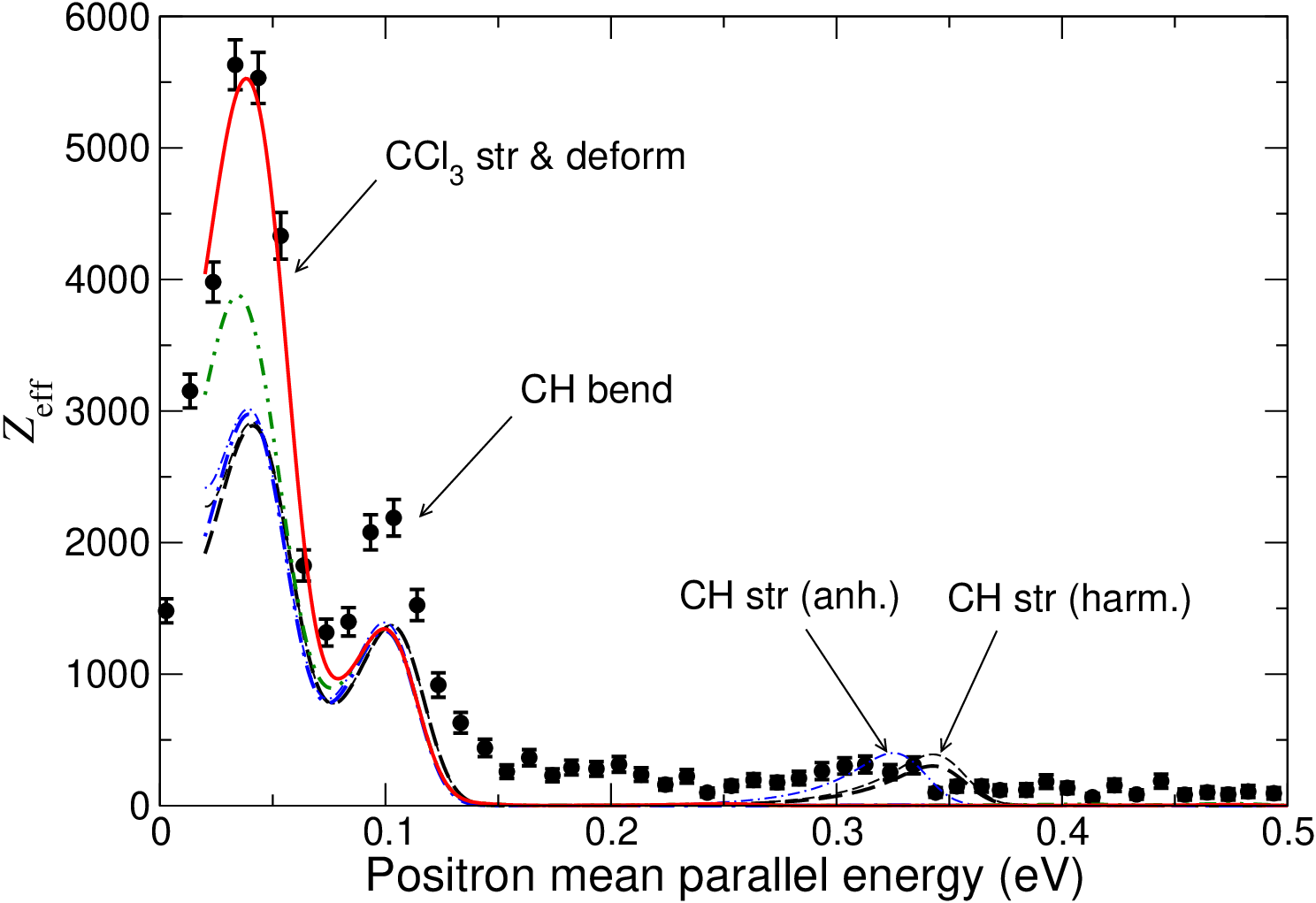}
\caption{Comparison between the calculated resonant $\Z$ and experimental data for chloroform ($\eps _b=40$~meV). Theoretical values from Eq.~(\ref{eq:Zeff}) include VFRs due to: modes (harmonic approximation), thick dashed line; same with $\Gamma _\nu^e/\Gamma _\nu =1$, thin dashed line; modes (anharmonic), thick dot-dashed line; same with $\Gamma _\nu^e/\Gamma _\nu =1$, thin dot-dashed line; 1- and 2-quantum excitations (anharmonic), dot-dot-dashed line; 1--3-quantum excitations (anharmonic), solid line. Solid circles show the experimental data from Refs. \cite{Jones2013,Natisin_thesis}.}
\label{fig:chcl3}
\end{figure}

In the simplest calculation, we use the vibrational data obtained in the harmonic approximation which allows only for single-quantum excitations of the modes in the sum in Eq.~(\ref{eq:Zeff}). This $\bar{Z}_{\rm eff}^{(\rm res)} (\epsilon )$ is shown by the thick dashed line in Fig.~\ref{fig:chcl3}, and the parameters of the corresponding VFR are listed in Table~\ref{tab:VFR_chcl3}. Their contributions to the $\Z$ spectrum are determined by the ratios $\Gamma _\nu^e/\Gamma _\nu $, and are also scaled with the resonance energy as $1/\sqrt{\eps _\nu }$. Shown by the thin dashed line is the maximum signal that could be produced by the mode-based resonances, in which we set $\Gamma _\nu^e/\Gamma _\nu =1$ for each of the resonances. Note that these two curves are almost indistinguishable except near the CH-stretch peak.

\begin{table*}[ht]
\caption{Parameters of vibrational Feshbach resonances for positron annihilation in chloroform. Resonance energies $\eps _\nu $ are in meV, resonance widths are in a.u. The notation $a[b]$ means $a\times 10^b$.}
\label{tab:VFR_chcl3}
\begin{ruledtabular}
\begin{tabular}{cccccccccccccc}
 & & & \multicolumn{3}{c}{Harmonic} & \multicolumn{4}{c}{Anharmonic, 1-quantum} &
 \multicolumn{4}{c}{Anharmonic, 1--3-quantum} \\
\cline{4-6} \cline{7-10} \cline{11-14}
VFR & Symm. & $g_\nu $ & $\eps _\nu $ & $\Gamma _\nu ^e$ & $\Gamma _\nu ^e/\Gamma _\nu $ & $\eps _\nu $ & $\Gamma _\nu ^e$ & $\Gamma _\nu ^v$ & $\Gamma _\nu ^e/\Gamma _\nu $ & $\eps _\nu $ & $\Gamma _\nu ^e$ & $\Gamma _\nu ^v$ & $\Gamma _\nu ^e/\Gamma _\nu $ \\
\hline
$\nu _3$ & $a_1$ & 1 & 6   & 2.21[-9] & 0.242 & 5   & 1.85[-9] & 1.85[-9] & 0.210 & 5  & 1.85[-9] & 1.85[-9] & 0.210 \\
$\nu _2$ & $a_1$ & 1 & 44  & 2.95[-7] & 0.977 & 43  & 3.07[-7] & 3.09[-7] & 0.973 & 43  & 3.07[-7] & 3.09[-7] & 0.973 \\
$\nu _5$ & $e$   & 2 & 58  & 1.01[-5] & 0.999 & 56  & 9.90[-6] & 9.91[-6] & 0.998 & 56  & 9.90[-6] & 9.91[-6] & 0.998 \\
$\nu _4$ & $e$   & 2 & 114 & 2.23[-6] & 0.997 & 111 & 2.01[-6] & 2.10[-6] & 0.954 & 111 & 2.01[-6] & 2.11[-6] & 0.953 \\
$\nu _1$ & $a_1$ & 1 & 354 & 2.38[-8] & 0.774 & 337 & 1.93[-9] & 1.72[-7] & 0.011 & 337 & 1.93[-9] & 1.74[-6] & 0.001\\
$\nu_3+\nu_6$ & $e$ & 2 & -- & -- & -- & -- & -- & -- & -- & 37 & 6.78[-9] & 8.58[-9] & 0.437 \\
$2\nu_3$ & $a_1$ & 1 & -- & -- & -- & -- & -- & -- & -- & 50 & 7.02[-10] & 4.36[-9] & 0.062 \\
$\nu_2+\nu_6$ & $e$ & 2 & -- & -- & -- & -- & -- & -- & -- & 75 & 3.11[-8] & 3.39[-7] & 0.090 \\
$3\nu_6$ & $e$ & 2 & -- & -- & -- & -- & -- & -- & -- & 56 & 1.01[-8] & 1.01[-8] & 0.592 \\
$3\nu_6$ & $a_1$ & 1 & -- & -- & -- & -- & -- & -- & -- & 56 & 1.24[-9] & 1.26[-9] & 0.151 \\
$3\nu_6$ & $a_1$ & 1 & -- & -- & -- & -- & -- & -- & -- & 56 & 8.44[-9] & 8.48[-9] & 0.547
\end{tabular}
\end{ruledtabular}
\end{table*}

For the lowest energy mode $\nu _6$, we have $\omega _{\nu0}<\eps _b$, and its ``resonance'' lies below threshold (i.e., at negative positron energies) and does not contribute to $\Z$. Of the remaining modes, the five vibrational states corresponding to $\nu _2$, $\nu_4$ and $\nu_5$ excitations have $\Gamma _\nu^e/\Gamma _\nu \approx 1$, and give maximum contributions to $\Z$. This a consequence of a sufficiently strong dipole coupling of these excited states to the vibrational ground state, such that $\Gamma _\nu ^e\gg \Gamma ^a$, and
\begin{equation}\label{eq:GG1}
\frac{\Gamma _\nu^e}{\Gamma _\nu }=\frac{\Gamma _\nu^e}{\Gamma _\nu^e+\Gamma ^a}\simeq 1,
\end{equation}
[cf. Eq.~(\ref{eq:Gam_nu}) and note that $\Gamma _\nu^v=\Gamma _\nu^e$ in the harmonic approximation]. The other two modes, $\nu _3$ and $\nu _1$, have the smallest dipole transition amplitudes (see Table~\ref{tab:chcl3}), which results in small elastic widths $\Gamma _\nu^e$ that are comparable to $\Gamma ^a$. Here the elastic width of the $\nu_3$ resonance is further suppressed due to its low energy (6 meV), since $\Gamma _\nu^e\propto k_\nu^3$ for $\eps_\nu \ll \omega _{\nu 0}$ [see Eq.~(\ref{eq:Gnu0})]. This gives the ratio $\Gamma _\nu^e/\Gamma _\nu =0.242$ and 0.774, for $\nu _3$ and $\nu _1$, respectively.

Compared with experiment, the annihilation rate obtained in the harmonic approximation reproduces the positions of the two main $\Z$ peaks, at $\epsilon =0.04$ and 0.1~eV, but underestimates their magnitudes by as much as a factor of two (for the low-energy peak).

Including anharmonic corrections has a relatively small effect on the dipole strengths of all modes except $\nu_1$, for which it decreases by a factor of ten (see Table~\ref{tab:chcl3}). It is thus natural that, apart from a small downshift of the resonance energies, the anharmonic calculation which accounts only for mode-based VFRs (thick solid line in Fig.~\ref{fig:chcl3}) gives $\Z$ in close agreements with the harmonic result. The exception here is the CH-stretch peak expected near 0.33~eV. Its contribution can only be seen in $\bar{Z}_{\rm eff}^{(\rm res)} (\epsilon )$ when in which we artificially set $\Gamma _\nu^e/\Gamma _\nu =1$ (thin dot-dashed line), while the calculated contribution of this resonances is determined by $\Gamma _\nu^e/\Gamma _\nu =0.011$.

Such a strong reduction of the width ratio cannot be explained by the decreased magnitude of the elastic width $\Gamma _\nu^e$. In fact, the quenching of the $\nu_1$ resonance is due to strong vibrationally inelastic escape from this VFR. This can be seen from the value of the positron escape width $\Gamma _\nu^v=1.72\times 10^{-7}$~a.u., which is two order of magnitude greater than $\Gamma _\nu^e=1.93\times 10^{-9}$~a.u., making
\[
\frac{\Gamma _\nu^e}{\Gamma _\nu }=\frac{\Gamma _\nu^e}{\Gamma _\nu^v+\Gamma ^a}\ll 1.
\]
For a calculation that includes only single-quantum vibrational excitations, the large vibrationally inelastic escape is due to anharmonic coupling between the modes. A detailed analysis of the escape width shows that the largest contribution to it comes from the $\nu_1\rightarrow \nu_4$ transition, which gives 87\% of the width $\Gamma _\nu^v$. The anharmonic coupling between the CH-stretch ($\nu _1$) and CH-bend ($\nu_4$) was investigated earlier by observing the $\nu _1+\nu_4$ combination band in near-infrared absorption and supported by density-functional calculations \cite{Nishida2012}. Suppression of the CH-stretch VFR due to inelastic escape was also inferred empirically in previous positron annihilation studies of chloroform and chloroform-$d_1$ \cite{Jones2013,Danielson2013} and in fluoroalkanes \cite{Natisin_thesis,RMP2010,YS08}.

Including 2- and 3-quantum vibrational excitations in the calculation of $\bar{Z}_{\rm eff}^{(\rm res)}$ has a dramatic effect on the low-energy $\Z$ peak, almost doubling its magnitude (see dot-dot-dashed and solid lines in Fig.~\ref{fig:chcl3}). The leading 2- and 3-quantum contributions are listed in Table~\ref{tab:VFR_chcl3}. Nearly all of them involve the CCl$_3$ asymmetric deformation mode $\nu_6$, whose 1-quantum excitation lies below threshold. The dipole strength of the $\nu_6$ mode itself is rather small. However, the energies of $3\nu_6$ overtones ($\approx 776~\text{cm}^{-1}$) lie very close to the strongest infrared-active mode $\nu_5$ (772~cm$^{-1}$), which may explain the origins of their dipole strengths. The calculated $\Z$ is now in much better agreement with experiment, though it still underestimates the height of the peak at 0.1~eV. The present $\bar{Z}_{\rm eff}^{(\rm res)}$ also cannot account for the observed annihilation rates at $\eps >0.15$~eV, which do not display any obvious resonant features. This discrepancy will be discussed in Sec.~\ref{sec:disc}.

\subsection{Chloroform-$d_1$}

It is interesting to compare the annihilation rate for chloroform with that of its deuterated analog, chloroform-$d_1$ (CDCl$_3$). Deuteration of a molecule has only a small effect on its electronic properties. In particular, the positron binding energies for CHCl$_3$ and CDCl$_3$ can be taken to be the same \cite{Jones2013}, so we use $\eps _b=40$~meV (see also Ref.~\cite{YGLS08} for methyl halides CH$_3$X and CD$_3$X).

Table \ref{tab:cdcl3} shows the mode frequencies and dipole transition strengths for chloroform-$d_1$. As in the case of chloroform, including the anharmonic corrections has a greater effect on higher-frequency modes, in particular, CD-stretch ($\nu _1$) and CD-bend ($\nu _4$). Due to the larger mass of the deuterium atom, the frequencies of both of these modes are noticeably lower than in chloroform. The calculated dipole strengths are generally in accord with infrared-intensity data.

\begin{table}[ht]
\caption{Vibrational mode energies and dipole transition strengths 
$g_\nu D_{0\nu }^2$ ($g_\nu$ being the mode degeneracy) for chloroform-$d_1$. The notation $a[b]$ means $a\times 10^b$.}
\label{tab:cdcl3}
\begin{ruledtabular}
\begin{tabular}{ccrrrccc}
\multicolumn{2}{l}{Mode and} & \multicolumn{3}{c}{$\omega $ (cm$^{-1}$)} & \multicolumn{3}{c}{$g_\nu D_{0\nu }^2$ (a.u.)} \\
\cline{3-5} \cline{6-8} 
\multicolumn{2}{l}{symmetry} & Harm. & Anh. & Sel.\footnote{Selected values from NIST tables~\cite{NIST}.} & Harm. & Anh. &
Exp.\footnote{Values derived from integrated intensity data \cite{Bishop1982}.}\\
\hline
$\nu _1$ & $a_1$ & 2341 & 2268 & 2266 & 1.94[-7] & 3.92[-6] & 2.95[-6]\\
$\nu _2$ & $a_1$ &  658 &  651 &  659 & 3.90[-4] & 4.12[-4] & 4.38[-4]\\
$\nu _3$ & $a_1$ &  366 &  363 &  369 & 5.40[-5] & 5.26[-5] & 7.40[-5]\\
$\nu _4$ & $e$   &  931 &  914 &  914 & 8.32[-3] & 7.76[-3] & 4.70[-3]\\
$\nu _5$ & $e$   &  760 &  745 &  749 & 1.37[-2] & 1.45[-2] & 1.74[-2]\\
$\nu _6$ & $e$   &  261 &  259 &  262 & 3.30[-5] & 2.58[-5] & 3.71[-5]\\
\end{tabular}
\end{ruledtabular}
\end{table}

\begin{figure}
\includegraphics*[width=0.48\textwidth]{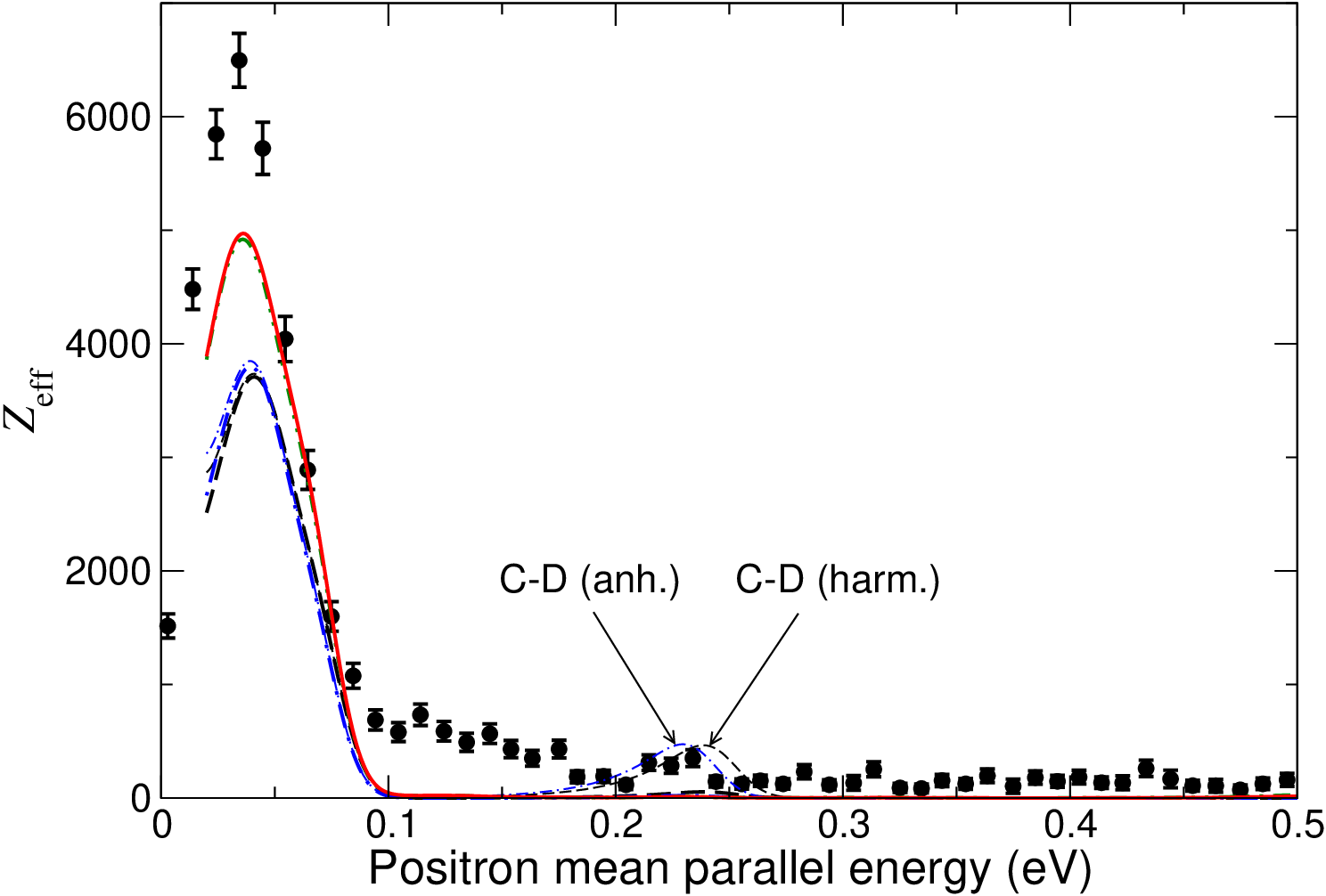}
\caption{Comparison between the calculated resonant $\Z$ and experimental data for chloroform-$d_1$ ($\eps _b=40$~meV). Theoretical values from Eq.~(\ref{eq:Zeff}) include VFRs due to: modes (harmonic approximation), thick dashed line; same with $\Gamma _\nu^e/\Gamma _\nu =1$, thin dashed line; modes (anharmonic), dot-dashed line; same with $\Gamma _\nu^e/\Gamma _\nu =1$, thin dot-dashed line; 1- and 2-quantum excitations (anharmonic), dot-dot-dashed line; 1--3-quantum excitations (anharmonic), solid line. Solid circles show the experimental data from Refs. \cite{Jones2013,Natisin_thesis}.}
\label{fig:cdcl3}
\end{figure}

Figure \ref{fig:cdcl3} shows the resonant annihilation rate $\bar{Z}_{\rm eff}^{(\rm res)}$ calculated in the harmonic approximation and with anharmonic corrections, in comparsion with the measured $\Z$ \cite{Jones2013,Natisin_thesis}. Parameters of the VFRs which give the dominant contribution to $\bar{Z}_{\rm eff}^{(\rm res)}$ are listed in Table~\ref{tab:VFR_cdcl3}. The frequency of the CD-bend ($\nu_4$) in chloroform-$d_1$ is close to that of degenerate CCl$_3$-stretch mode ($\nu_5$). This leads to disappearance of the two-peak spectral shape of $\Z$ that was seen in chloroform, as the $\nu_4$ and lower-lying VFR cannot be resolved with a room-temperature trap-based positron beam.

\begin{table*}[ht]
\caption{Parameters of vibrational Feshbach resonances for positron annihilation in chloroform-$d_1$. Resonance energies $\eps _\nu $ are in meV, resonance widths are in a.u. The notation $a[b]$ means $a\times 10^b$.}
\label{tab:VFR_cdcl3}
\begin{ruledtabular}
\begin{tabular}{cccccccccccccc}
 & & & \multicolumn{3}{c}{Harmonic} & \multicolumn{4}{c}{Anharmonic, 1-quantum} &
 \multicolumn{4}{c}{Anharmonic, 1--3-quantum} \\
\cline{4-6} \cline{7-10} \cline{11-14}
VFR & Symm. & $g_\nu $ & $\eps _\nu $ & $\Gamma _\nu ^e$ & $\Gamma _\nu ^e/\Gamma _\nu $ & $\eps _\nu $ & $\Gamma _\nu ^e$ & $\Gamma _\nu ^v$ & $\Gamma _\nu ^e/\Gamma _\nu $ & $\eps _\nu $ & $\Gamma _\nu ^e$ & $\Gamma _\nu ^v$ & $\Gamma _\nu ^e/\Gamma _\nu $ \\
\hline
$\nu _3$ & $a_1$ & 1 & 5   & 2.23[-9] & 0.243 & 5   & 1.92[-9] & 1.92[-9] & 0.217 & 5  & 1.92[-9] & 1.92[-9] & 0.217 \\
$\nu _2$ & $a_1$ & 1 & 42  & 2.62[-7] & 0.974 & 41  & 2.69[-7] & 2.70[-7] & 0.971 & 41  & 2.69[-7] & 2.70[-7] & 0.971  \\
$\nu _5$ & $e$   & 2 & 54  & 6.35[-6] & 0.999 & 52  & 6.44[-6] & 6.44[-6] & 0.998 & 52  & 6.44[-6] & 6.44[-6] & 0.998 \\
$\nu _4$ & $e$   & 2 & 76 & 5.68[-6] & 0.997 & 73 & 5.14[-6] & 5.23[-6] & 0.981 & 73 & 5.14[-6] & 5.23[-6] & 0.981 \\
$\nu _1$ & $a_1$ & 1 & 250 & 9.13[-10] & 0.116 & 241 & 1.79[-8] & 4.01[-7] & 0.044 & 241 & 1.79[-8] & 8.06[-7] & 0.022 \\
$\nu_3+\nu_6$ & $e$ & 2 & -- & -- & -- & -- & -- & -- & -- & 37 & 6.71[-9] & 8.58[-9] & 0.432 \\
$2\nu_3$ & $a_1$ & 1 & -- & -- & -- & -- & -- & -- & -- & 50 & 2.92[-10] & 4.11[-9] & 0.026 \\
$\nu_2+\nu_6$ & $e$ & 2 & -- & -- & -- & -- & -- & -- & -- & 73 & 1.84[-8] & 4.41[-7] & 0.411 \\
$\nu_3+2\nu_6$ & $a_1$ & 1 & -- & -- & -- & -- & -- & -- & -- & 69 & 7.91[-10] & 1.60[-8] & 0.034 \\
$2\nu_3+\nu_6$ & $e$ & 2 & -- & -- & -- & -- & -- & -- & -- & 82 & 2.65[-10] & 1.77[-8] & 0.011 \\
\end{tabular}
\end{ruledtabular}
\end{table*}

The $\Z$ due to single-mode (1-quantum) VFR in the harmonic and anharmonic calculations are similar. As in the case of chloroform, the lowest resonance ($\nu _3$ at 5~meV) and highest resonance ($\nu_1$ at 241~meV) are suppressed, with $\Gamma _\nu^e/\Gamma _\nu =0.243$ and 0.116 (harmonic), and 0.217 and 0.044 (anharmonic), respectively. The ratio $\Gamma _\nu^e/\Gamma _\nu $ is suppressed due to coupling between the CD-stretch and CD-bend modes, which enables vibrationally inelastic escape. It is further suppressed when 2- and 3-quantum excitations are included, with $\nu_1$ VFR decaying effectively into $\nu _2+\nu_4$ and $\nu_2+\nu_5$ final states.

Including 2- and 3-quantum excitations increases the main $\Z$ peak (see Fig.~\ref{fig:cdcl3}), although their effect is not as large as in normal chloroform. Only two combination vibrations ($\nu_3+\nu_6$ and $\nu_2+\nu_6$) give sizeable contriubutions, while the remaining multiquantum resonances have $\Gamma _\nu^e/\Gamma _\nu $ values of a few per cent or less. The magnitude of the peak in the calculated $\Z$ remains about 25\% smaller than that from the measurements. The calculation also fails to account for the signal above 0.1~eV (see Sec.~\ref{sec:disc}).

\subsection{1,1-Dichloroethylene}

The next molecule considered is the six-atom 1,1-dichloroethylene (C$_2$H$_2$Cl$_2$). It belongs to the $C_{2v}$ point group and has 12 nondegenerate vibrational modes of four symmetry species: $a_1$, $a_2$, $b_1$ and $b_2$, of which all except $a_2$ ($\nu _6$ mode) are infrared-active. The mode frequencies and dipole strengths are listed in Table~\ref{tab:c2h2cl2}.

\begin{table}[ht]
\caption{Vibrational mode energies and dipole transition strengths 
$D_{0\nu }^2$ for 1,1-dichloroethylene. The notation $a[b]$ means $a\times 10^b$.}
\label{tab:c2h2cl2}
\begin{ruledtabular}
\begin{tabular}{cclrrrcc}
\multicolumn{3}{c}{Mode symmetry} & \multicolumn{3}{c}{$\omega $ (cm$^{-1}$)} & \multicolumn{2}{c}{$D_{0\nu }^2$ (a.u.)} \\
\cline{4-6} \cline{7-8} 
\multicolumn{3}{c}{and type \cite{NIST}} & Harm. & Anh. & Sel.\footnote{Selected values from NIST tables~\cite{NIST}.} & \multicolumn{1}{c}{Harm.} & \multicolumn{1}{c}{Anh.} \\
\hline
$\nu _1$   & $a_1$ & CH$_2$ s-str  & 3179 & 3044 & 3035 & 1.95[-5] & 7.62[-6]\footnote{Sum of $\nu_2+\nu_3$ (3031~cm$^{-1}$, 75\%) and $\nu_1$ (3044~cm$^{-1}$, 25\%).} \\
$\nu _2$   & $a_1$ & CC str        & 1647 & 1622 & 1627 & 2.11[-3] & 2.40[-3]\footnote{Sum of $2\nu_9$ (1561~cm$^{-1}$, 36\%) and $\nu_2$ (1622~cm$^{-1}$, 64\%).} \\
$\nu _3$   & $a_1$ & CH$_2$ scis   & 1402 & 1422 & 1400 & 8.46[-7] & 2.24[-5] \\
$\nu _4$   & $a_1$ & CCl$_2$ s-str &  607 &  599 &  603 & 1.28[-3] & 1.31[-3] \\
$\nu _5$   & $a_1$ & CCl$_2$ scis  &  298 &  296 &  299 & 7.51[-6] & 8.24[-6] \\
$\nu _6$   & $a_2$ & Torsion       &  696 &  680 &  686 & 0 & 0 \\
$\nu _7$   & $b_1$ & CH$_2$ a-str  & 3282 & 3136 & 3130 & 1.12[-6] & 8.08[-7]\footnote{Sum of $\nu_7$ (3136~cm$^{-1}$, 87\%) and $\nu_2+\nu_6+\nu_{11}$ (3143~cm$^{-1}$, 13\%).} \\
$\nu _8$   & $b_1$ & CH$_2$ rock   & 1112 & 1091 & 1095 & 4.62[-3] & 5.36[-3]\footnote{Sum of $\nu_5+\nu_9$ (1081~cm$^{-1}$, 46\%), $\nu_8$ (1091~cm$^{-1}$, 39\%), 3$\nu_{10}$ (1116~cm$^{-1}$, 5\%), and $\nu_6+\nu_{12}$ (1137~cm$^{-1}$, 10\%).} \\
$\nu _9$   & $b_1$ & CCl$_2$ a-str &  806 &  791 &  800 & 5.49[-3] & 5.70[-3] \\
$\nu _{10}$& $b_1$ & CCl$_2$ rock  &  372 &  371 &  372 & 8.61[-5] & 6.84[-5] \\
$\nu _{11}$& $b_2$ & CH$_2$ wag    &  882 &  865 &  875 & 3.04[-3] & 3.00[-3] \\
$\nu _{12}$& $b_2$ & CCl$_2$ wag   &  460 &  455 &  460 & 5.75[-4] & 1.99[-5] \\
\end{tabular}
\end{ruledtabular}
\end{table}

\begin{figure}[ht]
\includegraphics*[width=0.48\textwidth]{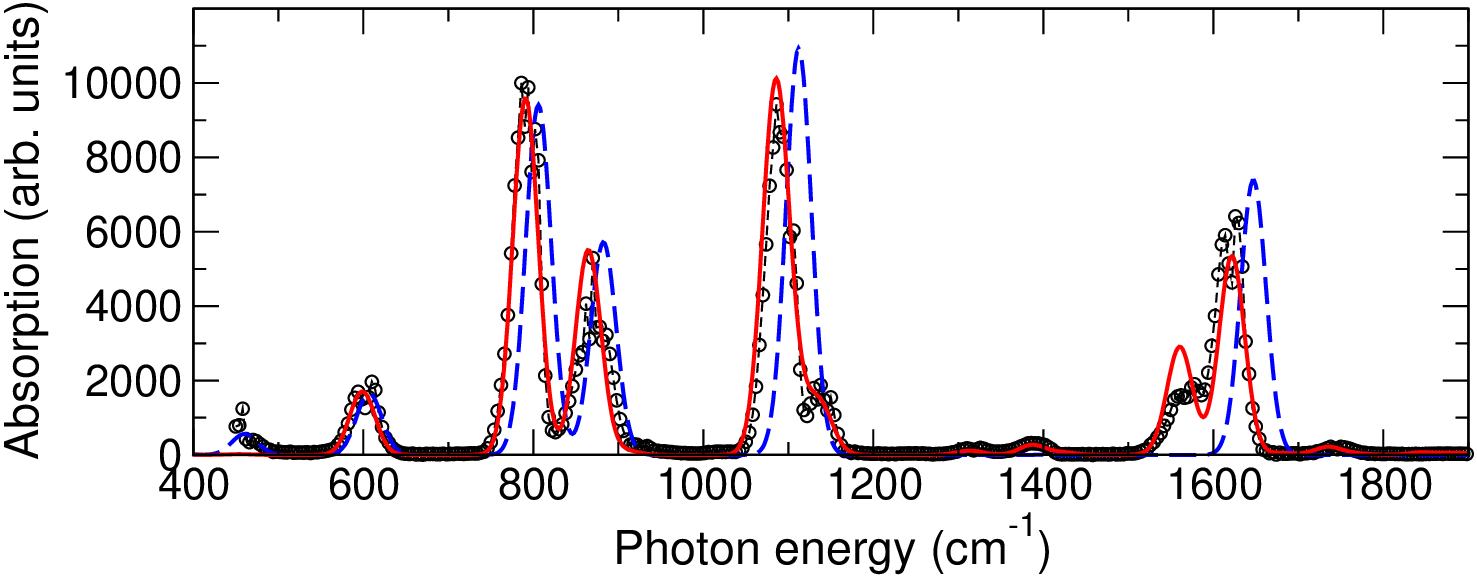}
\caption{Calculated and measured \cite{NIST} infrared absorption intensity for 1,1-dichloroethylene: harmonic approximation, long-dashed line; anharmonic with 1--3-quantum excitations, solid line; experiment, circles connected by short-dashed line
(see text for details).}
\label{fig:c2h2cl2_abs}
\end{figure}

The anharmonic calculation includes all 1-, 2-, and 3-quantum excitations of the 12 modes, i.e., a total of 455 states (including the ground state). Table~\ref{tab:c2h2cl2} shows that anharmonic corrections have the largest effect on the frequencies of the CH$_2$-stretch modes, bringing them into close agreement with experiment. Since experimental data are not available for the strengths of individual modes, we compute the infrared absorption intensity, which is proportional to $\sum _\nu \omega _{\nu 0}|D_{\nu 0}|^2\delta(\omega -\omega _{\nu 0})$. A comparison between the calculated and measured absorption intensities is shown in Fig.~\ref{fig:c2h2cl2_abs}, for which we broaden each absorption line by a Gaussian with FWHM of 35~cm$^{-1}$. It is clear that anharmonic effects lead to a much better agreement with experiment. The anharmonic calculation also shows prominent contributions of overtones and combination vibrations, e.g., $2\nu _9$ at 1561~cm$^{-1}$ or $\nu_6+\nu_{12}$ at 1137~cm$^{-1}$. When these contributions are near-resonant with the modes and of the same symmetry, we have included them in the total dipole strength values in Table~\ref{tab:c2h2cl2} (last column), with details provided in the footnotes. In some cases the contributions of the modes and overtones or combinations are of comparable strength, which makes identification of such modes ambiguous.

\begin{figure}[ht]
\includegraphics*[width=0.48\textwidth]{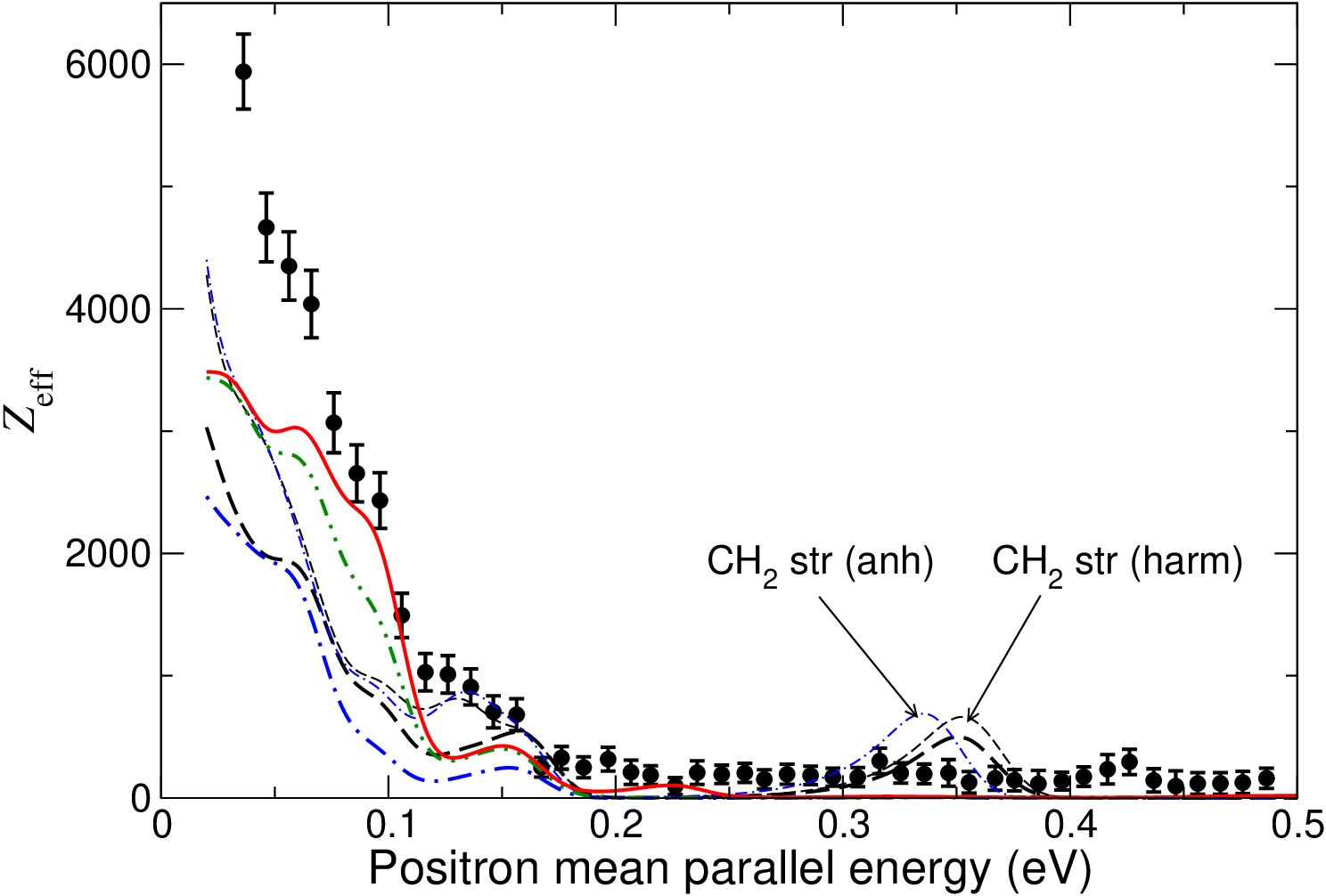}
\caption{Comparison between the calculated resonant $\Z$ and experimental data for 1,1-dichloroethylene ($\eps _b=35$~meV). Theoretical values from Eq.~(\ref{eq:Zeff}) include VFRs due to: modes (harmonic approximation), thick dashed line; same with $\Gamma _\nu^e/\Gamma _\nu =1$, thin dashed line; modes (anharmonic), dot-dashed line; same with $\Gamma _\nu^e/\Gamma _\nu =1$, thin dot-dashed line; 1- and 2-quantum excitations (anharmonic), dot-dot-dashed line; 1--3-quantum excitations (anharmonic), solid line. Solid circles show the experimental data from Ref. \cite{Natisin_thesis}.}
\label{fig:c2h2cl2}
\end{figure}

Figure \ref{fig:c2h2cl2} shows values of $\bar{Z}_{\rm eff}^{(\rm res)}$ for 1,1-dichloroethylene obtained from Eq.~(\ref{eq:Zeff}) using harmonic and anharmonic vibrational data, in comparison with measured $\Z$ \cite{Natisin_thesis}.
The expermental $\Z$ spectrum does not display any clearly resolved and unambiguously assignable VFR peaks that would enable one to determine the binding energy. Hence, we use $\eps _b=35$~meV obtained by fitting the measured $\Z$ with beam-energy-distribution-broadened resonances of the modes with adjustable vertical scaling \cite{Natisin_thesis}. The corresponding annihilation width is $\Gamma ^a=6.5\times 10^{-9}$~a.u.

In the harmonic approximation, seven modes give contributions close to the theoretical maximum, $\Gamma _\nu ^e/\Gamma _\nu \approx 1$ (see VFR parameters in Table~\ref{tab:VFR_c2h2cl2}). The low-lying $\nu _5$ and $\nu_{10}$ are suppressed because of the small incident positron energy, while $\nu _3$ and $\nu _7$ have small dipole strengths; the $\nu _6$ mode is infrared-inactive, hence it does not contribute at all.

When anharmonic corrections are taken into account in the calculation of $\bar{Z}_{\rm eff}^{(\rm res)}$, which includes only 1-quantum excitations (thick dot-dashed line in Fig.~\ref{fig:c2h2cl2}), only three resonances retain $\Gamma _\nu ^e/\Gamma _\nu \approx 1$. Compared with the harmonic calculation, the contribution of the $\nu_{12}$ VFR is reduced by a factor of two because of the reduction in its dipole strength (see Table~\ref{tab:c2h2cl2}). Values of $\Gamma _\nu ^e/\Gamma _\nu $ for the $\nu _8$ and $\nu _2$ VFRs are also halved, this time due to coupling between these modes and $\nu_5$ and $\nu_9$, respectively. Finally, the most dramatic effect is the total quenching of the CH-stretch peaks at 0.35~eV. For both $\nu_1$ and $\nu_7$ modes, there is a large dipole coupling with $\nu_2$, $\nu_3$, and $\nu_{11}$, with $\nu_7$ also coupled strongly to $\nu_8$. As a result, the corresponding VFRs decay by vibrationally inelastic escape ($\Gamma _\nu^v\approx 10^2\Gamma _\nu^e$), and so they do not produce a noticeable contribution to the $\Z$ spectrum. They can only be seen in the calculation in which we set $\Gamma _\nu ^e/\Gamma _\nu =1$ (thin dot-dashed line in Fig.~\ref{fig:c2h2cl2}).

\begin{table*}[ht]
\caption{Parameters of vibrational Feshbach resonances for positron annihilation in 1,1-dichloroethylene. Resonance energies $\eps _\nu $ are in meV, resonance widths are in a.u. The notation $a[b]$ means $a\times 10^b$.}
\label{tab:VFR_c2h2cl2}
\begin{ruledtabular}
\begin{tabular}{ccccccccccccc}
 &  & \multicolumn{3}{c}{Harmonic} & \multicolumn{4}{c}{Anharmonic, 1-quantum} &
 \multicolumn{4}{c}{Anharmonic, 1--3-quantum} \\
\cline{3-5} \cline{6-9} \cline{10-13}
VFR & Symm. & $\eps _\nu $ & $\Gamma _\nu ^e$ & $\Gamma _\nu ^e/\Gamma _\nu $ & $\eps _\nu $ & $\Gamma _\nu ^e$ & $\Gamma _\nu ^v$ & $\Gamma _\nu ^e/\Gamma _\nu $ & $\eps _\nu $ & $\Gamma _\nu ^e$ & $\Gamma _\nu ^v$ & $\Gamma _\nu ^e/\Gamma _\nu $ \\
\hline
$\nu _5$ & $a_1$ & 2  & 7.46[-11] & 0.011 & 2   & 6.91[-11] & 6.91[-11] & 0.011 & 2   & 6.91[-11] & 6.91[-11] & 0.011 \\
$\nu _{10}$ & $b_1$ & 11  & 1.05[-8] & 0.617 & 11  & 8.19[-9] & 8.19[-9] & 0.557 & 11  & 8.19[-9] & 8.19[-9] & 0.557  \\
$\nu _{12}$ & $b_2$   & 22  & 1.78[-7] & 0.965 & 21  & 5.89[-9] & 5.89[-9] & 0.475 & 21  & 5.89[-9] & 5.89[-9] & 0.475  \\
$\nu _4$ & $a_1$   & 40 & 8.52[-7] & 0.992 & 39 & 8.47[-7] & 8.47[-7] & 0.992 & 39 & 8.47[-7] & 8.47[-7] & 0.992  \\
$\nu _9$ & $b_1$ & 65 & 6.44[-6] & 0.999 & 63 & 6.47[-6] & 6.47[-6] & 0.999 & 63 & 6.47[-6] & 6.47[-6] & 0.999 \\
$\nu_{11}$ & $b_2$ & 74 & 4.17[-6] & 0.998 & 72 & 3.96[-6] & 3.97[-6] & 0.997 & 72 & 3.96[-6] & 3.97[-6] & 0.997 \\
$\nu_8$ & $b_1$ & 103 & 9.00[-6] & 0.999 & 100 & 3.96[-6] & 7.13[-6] & 0.556 & 100 & 3.96[-6] & 7.13[-6] & 0.556 \\
$\nu_3$ & $a_1$ & 139 & 2.25[-9] & 0.257 & 141 & 6.05[-8] & 7.98[-7] & 0.075 & 141 & 6.05[-8] & 8.02[-7] & 0.075 \\
$\nu_2$ & $a_1$ & 169 & 6.81[-6] & 0.999 & 166 & 4.87[-6] & 1.09[-5] & 0.448 & 166 & 4.87[-6] & 1.09[-5] & 0.447 \\
$\nu_1$ & $a_1$ & 359 & 1.25[-7] & 0.951 & 342 & 1.17[-8] & 1.42[-6] & 0.008 & 342 & 1.17[-8] & 7.06[-6] & 0.002 \\
$\nu_7$ & $b_1$ & 372 & 7.41[-9] & 0.533 & 354 & 4.46[-9] & 3.94[-7] & 0.011 & 354 & 4.46[-9] & 1.21[-6] & 0.004 \\
$2\nu _5$ & $a_1$ &--&--&--&--&--&--&--& 39 & 1.73[-8] & 1.74[-8] & 0.723 \\
$2\nu _{12}$ & $a_1$ &--&--&--&--&--&--&--& 78 & 7.90[-8] & 9.23[-8] & 0.800 \\
$\nu _5+\nu_9$ & $b_1$ &--&--&--&--&--&--&--& 99 & 4.66[-6] & 8.39[-6] & 0.554 \\
$3\nu _{10}$ & $b_1$ &--&--&--&--&--&--&--& 103 & 5.18[-7] & 5.42[-7] & 0.943 \\
$\nu_6+\nu _{12}$ & $b_1$ &--&--&--&--&--&--&--& 106 & 1.04[-6] & 1.06[-6] & 0.975 \\
$2\nu _9$ & $a_1$ &--&--&--&--&--&--&--& 159 & 2.63[-6] & 1.07[-7] & 0.246 \\
$\nu_{11}+\nu _{12}$ & $a_1$ &--&--&--&--&--&--&--& 236 & 3.36[-6] & 1.52[-5] & 0.221 \\
\end{tabular}
\end{ruledtabular}
\end{table*}

Adding the contributions of 2- and 3-quantum resonances increases the calculated $\Z$ 
greatly compared with that which only includes VFRs of the modes, at energies below 0.2~eV. Parameters of the strongest 2- and 3-quantum VFRs are listed in Table~\ref{tab:VFR_c2h2cl2}. Besides these VFR, there are eight resonances with $\Gamma _\nu ^e/\Gamma _\nu $ in the range 0.01--0.05, and many resonances with smaller contributions, which only increase $\Z$ by few per cent. While the final calculated $\Z$ (solid line in Fig.~\ref{fig:c2h2cl2}) is lower than the measured annihilation rate, its main features are described quite well qualitatively by the theory. In particular, the calculations reproduce the rapid onset of the signal below 0.12 eV. Above 0.2~eV,
apart from the 2-quantum VFR $\nu_{11}+\nu _{12}$ at 236~meV, the calculated $\bar{Z}_{\rm eff}^{(\rm res)}$ is given by many small contributions of multiquantum resonances. Here, as in the case of chloroform, the calculation does not explain the observed magnitude of the annihilation rate (see Sec.~\ref{sec:disc}).

\subsection{Methanol}

Unlike the molecules studied above, methanol (CH$_3$OH) has only one mode below 1000~cm$^{-1}$ (torsion, at 295~cm$^{-1}$), and has quite strong transitions to its CH stretching fundamental levels. As a result, its $\Z$ spectrum is quite different, with a prominent peak at 0.35~eV due to CH-stretch vibrations \cite{YGLS08,Jones2012a}.

Table \ref{tab:ch3oh} shows the energies and dipole strengths of the vibrational modes in methanol, calculated in the harmonic approximation and with anharmonic corrections. Values derived from integrated intensity data in liquid methanol \cite{Bertie1997} are shown for comparison, except for $\nu_1$ and $\nu_{12}$. The latter modes are strongly affected by the molecular environment, and for these modes, calculated values from Ref.~\cite{Florian1997} are shown.

Anharmonic corrections are quite large in methanol, especially for the OH- and CH-stretch modes. They bring the mode energies into close agreement with experiment for all the modes except $\nu_3$ and $\nu_{12}$. The energy of $\nu_3$ is poorly described here, as no serious effort was made (apart from the deperturbation mentioned in Sec.~\ref{subsec:vib}) to treat the strong Fermi resonance that couples this level with the $2\nu_{10}$ overtone, and the torsional motion is not described well by the VPT2 model. However, these discrepancies are not expected to have a large effect on the calculated $\bar{Z}_{\rm eff}^{(\rm res)}$, in part, because averaging over the positron-beam energy distribution broadens the resonances to about 40~meV FWHM.

\begin{table*}[ht]
\caption{Vibrational mode energies and dipole transition strengths 
$D_{0\nu }^2$ for methanol. The notation $a[b]$ means $a\times 10^b$.}
\label{tab:ch3oh}
\begin{ruledtabular}
\begin{tabular}{cclrrrcccc}
\multicolumn{3}{c}{Mode symmetry} & \multicolumn{3}{c}{$\omega $ (cm$^{-1}$)} & \multicolumn{3}{c}{$D_{0\nu }^2$ (a.u.)} \\
\cline{4-6} \cline{7-9} 
\multicolumn{3}{c}{and type \cite{NIST}} & Harm. & Anh. & Sel.\footnote{Selected values from NIST tables~\cite{NIST}.} & Harm. & Anh. & Exp.\footnote{Values derived from integrated intensity data in liquid methanol \cite{Bertie1997},
except for $\nu_8$ \cite{Dangnhu1990}. The total intensity measured for $\nu_4$ and $\nu _{10}$ is split as 3:2 between the modes.
For $\nu_1$ and $\nu_{12}$ values from \textit{ab initio} CCSD(T)/6-311G(3df,2p) calculations \cite{Florian1997} are shown.} \\
\hline
$\nu _1$   & $a'$ & OH str          & 3865 & 3681 & 3681 & 4.17[-4] & 3.59[-4] & 4.77[-4] \\
$\nu _2$   & $a'$ & CH$_3$ d-str    & 3134 & 2989 & 3000 & 4.93[-4] & 5.64[-4] & 5.66[-4] \\
$\nu _3$   & $a'$ & CH$_3$ s-str    & 3013\footnotemark[4]\footnotetext[4]{The resonance between $\nu_3$ and $2\nu_{10}$ was ``deperturbed'' in the anharmonic calculation, by shifting their frequencies to 3015 and 1503~cm$^{-1}$.} & 2932 & 2844 & 1.04[-3] & 7.48[-3] & 5.13[-4]\\
$\nu _4$   & $a'$ & CH$_3$ d-deform & 1520 &  1477 & 1477 & 1.70[-4] & 1.52[-4] & 3.11[-4]\\
$\nu _5$   & $a'$ & CH$_3$ s-deform & 1484 &  1451 & 1455 & 1.15[-4] & 1.78[-5] & 1.31[-4]\\
$\nu _6$   & $a'$ & OH bend         & 1388 &  1337 & 1345 & 1.20[-3] & 1.06[-3]  & 9.12[-4]\\
$\nu _7$   & $a'$ & CH$_3$ rock     & 1090 &  1069 & 1060 & 1.24[-4] & 1.41[-4] & 6.95[-4] \\
$\nu _8$   & $a'$ & CO str          & 1060 & 1035 & 1033 & 6.18[-3] & 6.83[-3] & 5.92[-3]\\
$\nu _9$   & $a''$ & CH$_3$ d-str   & 3071 & 2935 & 2960 & 1.03[-3] & 9.83[-4] & 8.21[-4] \\
$\nu _{10}$& $a''$ & CH$_3$ d-deform& 1506\footnotemark[4] & 1464 & 1477 & 1.08[-4] & 1.03[-4] & 2.07[-4]\\
$\nu _{11}$& $a''$ & CH$_3$ rock    & 1181 & 1151 & 1165 & 4.11[-5] & 2.65[-5] & 7.62[-5]\\
$\nu _{12}$& $a''$ & torsion        &  304 &  249 &  295 & 2.17[-2] & 2.40[-2] & 2.45[-2]\\
\end{tabular}
\end{ruledtabular}
\end{table*}

In general, anharmonic effects change the dipole strengths of the modes by about 10\% (see Table~\ref{tab:ch3oh}). A larger effect is observed for $\nu_5$. This is likely related to some redistribution of the absorption strength in the range of $\nu_4$, $\nu _5$ and $\nu_6$ modes (which also includes the combination $\nu_{11}+\nu_{12}$ at 1407~cm$^{-1}$ with $D_{0\nu }^2=2.8\times 10^{-4}$~a.u.). The only anomaly in the anharmonic data is the large dipole strength of the $\nu_3$ mode, which is a consequence of incomplete deperturbation of its resonance with $2\nu_{10}$ at the harmonic level.

In the absence of gas-phase data for the absorption intensities (except \cite{Dangnhu1990} for the CO-stretch mode), the calculated transition dipole strengths can be compared with the values obtained in liquid methanol. The only vibrations that are strongly affected by the environment are the OH stretch (because of hydrogen bonding) and torsion, and so we use earlier theoretical data \cite{Florian1997} for these. For all modes (except $\nu_3$) there is a reasonable accord between the calculated and measured data. Larger discrepancies observed for the CH$_3$-rock and CH$_3$-d-deform modes can probably be attributed to the uncertainty in separating the intensities of the modes for overlapping bands. 

\begin{figure}[ht]
\includegraphics*[width=0.48\textwidth]{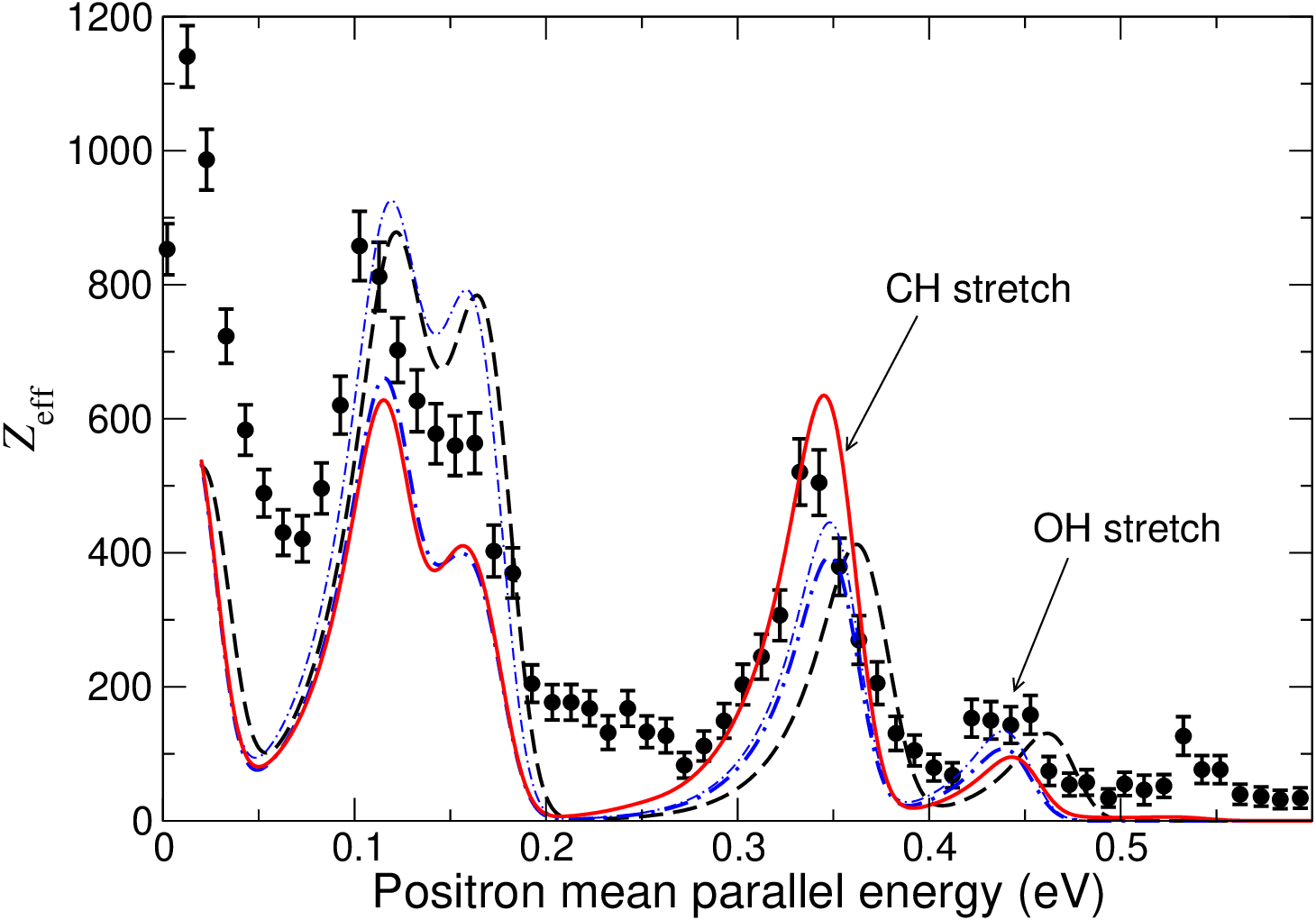}
\caption{Comparison between the calculated resonant $\Z$ and experimental data for methanol ($\eps _b=6$~meV). Theoretical values from Eq.~(\ref{eq:Zeff}) include VFRs due to: modes (harmonic approximation), thick dashed line; modes (anharmonic), dot-dashed line; same with $\Gamma _\nu^e/\Gamma _\nu =1$, thin dot-dashed line; 1--3-quantum excitations (anharmonic), solid line. Solid circles show the experimental data from Ref. \cite{Natisin_thesis}.}
\label{fig:ch3oh}
\end{figure}

Turning now to $\Z$, we use the binding energy $\eps _b=6$~meV \cite{Natisin_thesis}. This value and the measured annihilation rate shown in Fig.~\ref{fig:ch3oh}, correct the earlier experimental data \cite{YGLS08,Jones2012a}. The shape of the $\Z$ spectrum is similar to that observed in hydrocarbons \cite{Barnes2003}, with a broad feature between 0.1 and 0.2~eV due to CH$_3$ rocking and deformation, and a prominent CH-stretch peak at 0.35 eV, with the addition of a higher-energy OH-stretch peak at 0.43~eV.

In the harmonic approximation, all twelve VFRs of the modes contribute to $\bar{Z}_{\rm eff}^{(\rm res)}$ at the level close to the theoretical maximum $\Gamma _\nu ^e/\Gamma _\nu =1$ (see Table~\ref{tab:VFR_ch3oh}). This is a consequence of all modes having appreciable dipole coupling to the vibrational ground state and small annihilation width $\Gamma ^a=2.69\times 10^{-9}$~a.u. [cf. Eq.~(\ref{eq:GG1})]. Including the anharmonic effects lowers the mode energies and brings the positions of the $\Z$ maxima into closer agreement with experiment. Simultaneously, the ratio $\Gamma _\nu ^e/\Gamma _\nu $ drops for all the modes except $\nu_{12}$, especially in the range of CH$_3$ rocking and deformation modes. Here the $\bar{Z}_{\rm eff}^{(\rm res)}$ produced by the VFRs of the modes (thick dot-dashed line in Fig.~\ref{fig:ch3oh}) is markedly lower than that obtained by setting $\Gamma _\nu ^e/\Gamma _\nu =1$ (thin dot-dashed line) due to inelastic escape. The suppression of $\Gamma _\nu ^e/\Gamma _\nu $ is particularly large for $\nu_{5}$, which couples strongly to $\nu_{11}$. It is also significant for $\nu_6$ and $\nu_4$, which both couple to $\nu_{11}$, and for $\nu_{11}$ itself, which couples to all lower-lying modes ($\nu_7$, $\nu_8$ and $\nu_{12}$).

\begin{table*}[ht]
\caption{Parameters of vibrational Feshbach resonances for positron annihilation in methanol. Resonance energies $\eps _\nu $ are in meV, resonance widths are in a.u. The notation $a[b]$ means $a\times 10^b$.}
\label{tab:VFR_ch3oh}
\begin{ruledtabular}
\begin{tabular}{ccccccccccccc}
 &  & \multicolumn{3}{c}{Harmonic} & \multicolumn{4}{c}{Anharmonic, 1-quantum} &
 \multicolumn{4}{c}{Anharmonic, 1--3-quantum} \\
\cline{3-5} \cline{6-9} \cline{10-13}
VFR & Symm. & $\eps _\nu $ & $\Gamma _\nu ^e$ & $\Gamma _\nu ^e/\Gamma _\nu $ & $\eps _\nu $ & $\Gamma _\nu ^e$ & $\Gamma _\nu ^v$ & $\Gamma _\nu ^e/\Gamma _\nu $ & $\eps _\nu $ & $\Gamma _\nu ^e$ & $\Gamma _\nu ^v$ & $\Gamma _\nu ^e/\Gamma _\nu $ \\
\hline
$\nu _{12}$ & $a''$ & 32  & 1.31[-5] & 1.000 & 25   & 1.14[-5] & 1.14[-5] & 1.000 & 25   & 1.14[-5] & 1.14[-5] & 1.000  \\
$\nu _8$ & $a'$ & 126  & 1.20[-5] & 1.000 & 122  & 1.30[-5] & 1.30[-5] & 0.998 & 122  & 1.30[-5] & 1.30[-5] & 0.998  \\
$\nu _7$ & $a'$   & 129  & 2.46[-7] & 0.989 & 127  & 2.76[-7] & 2.88[-7] & 0.949 & 127  & 2.76[-7] & 2.90[-7] & 0.945  \\
$\nu _{11}$ & $a''$   & 140 & 8.68[-8] & 0.970 & 137 & 5.49[-8] & 1.74[-7] & 0.311 & 137 & 5.49[-8] & 3.76[-7] & 0.145  \\
$\nu _6$ & $a'$ & 166 & 2.87[-6] & 0.999 & 160 & 2.46[-6] & 3.47[-6] & 0.709 & 160 & 2.46[-6] & 3.53[-6] & 0.698 \\
$\nu_5$ & $a'$ & 178 & 2.89[-7] & 0.991 & 174 & 4.39[-8] & 2.77[-6] & 0.016 & 174 & 4.39[-8] & 2.81[-6] & 0.016 \\
$\nu_{10}$ & $a''$ & 181 & 2.75[-7] & 0.990 & 176 & 2.56[-7] & 2.71[-7] & 0.937 & 176 & 2.56[-7] & 2.71[-7] & 0.936 \\
$\nu_4$ & $a'$ & 182 & 4.33[-7] & 0.994 & 177 & 3.81[-7] & 9.77[-7] & 0.389 & 177 & 3.81[-7] & 9.83[-7] & 0.386 \\
$\nu_3$ & $a'$ & 368 & 4.29[-6] & 0.999 & 358 & 3.03[-5] & 3.51[-5] & 0.865 & 358 & 3.03[-5] & 6.87[-5] & 0.442 \\
$\nu_9$ & $a''$ & 375 & 4.33[-6] & 0.999 & 358 & 3.99[-6] & 4.33[-6] & 0.922 & 358 & 3.99[-6] & 4.99[-6] & 0.799 \\
$\nu_2$ & $a'$ & 383 & 2.09[-6] & 0.999 & 365 & 2.31[-6] & 2.66[-6] & 0.870 & 365 & 2.31[-6] & 2.83[-6] & 0.816 \\
$\nu_1$ & $a'$ & 473 & 2.03[-6] & 0.999 & 450 & 1.69[-6] & 2.10[-6] & 0.802 & 450 & 1.69[-6] & 3.89[-5] & 0.043 \\
$\nu_{11}+\nu_{12}$ & $a'$ &--&--&--&--&--&--&--& 168 & 6.76[-7] & 9.64[-6] & 0.070 \\
$\nu_5+\nu_{11}+\nu_{12}$ & $a'$ &--&--&--&--&--&--&--& 348 & 1.17[-6] & 1.82[-5] & 0.064 \\
$2\nu _5$ & $a'$ &--&--&--&--&--&--&--& 348 & 2.67[-6] & 8.15[-6] & 0.327 \\
$\nu_4+\nu_{11}+\nu_{12}$ & $a'$ &--&--&--&--&--&--&--& 350 & 1.14[-5] & 2.26[-7] & 0.507 \\
$2\nu _{10}$ & $a'$ &--&--&--&--&--&--&--& 351 & 1.92[-5] & 6.59[-5] & 0.291 \\
$\nu _5+\nu_{10}$ & $a''$ &--&--&--&--&--&--&--& 355 & 1.83[-7] & 3.32[-6] & 0.055 \\
$\nu_4+\nu _5$ & $a'$ &--&--&--&--&--&--&--& 356 & 2.72[-6] & 2.22[-5] & 0.123 \\
$\nu_4+\nu _{10}$ & $a''$ &--&--&--&--&--&--&--& 358 & 4.99[-7] & 1.63[-6] & 0.306 \\
$2\nu_4$ & $a'$ &--&--&--&--&--&--&--& 364 & 6.35[-6] & 8.60[-6] & 0.738 \\
$2\nu_6+\nu _7$ & $a'$ &--&--&--&--&--&--&--& 455 & 1.25[-5] & 1.91[-5] & 0.656 \\
\end{tabular}
\end{ruledtabular}
\end{table*}

Including the contributions of 2- and 3-quantum VFRs (solid line in Fig.~\ref{fig:ch3oh}) increases the height of the CH-stretch peak by about 50\% compared with that produced by 1-quantum excitations of the modes in the anharmonic approximation. It also moves it into the ``correct'' position, but makes only a small difference elsewhere. Parameters of VFRs of the modes and leading 2- and 3-quantum excitations with $\Gamma _\nu ^e/\Gamma _\nu >0.05$ are listed in the last four columns of Table~\ref{tab:VFR_ch3oh}. Multiquantum vibrational states provide additional inelastic escape channels which reduce the contributions of $\nu_3$ and $\nu_1$ single-quantum VFRs. However, their dominant effect is the emergence of additional, mostly 2-quantum,  VFRs, eight of which are in the range 348--364~meV. Their total contribution can be estimated as $\sum _\nu \Gamma _\nu ^e/\Gamma _\nu =2.4$, which is equivalent to ``two-and-a-half resonances'' contributing at the maximum level. The shape of both the CH$_3$-rocking and deformation part of the $\Z$ spectrum and the CH-stretch peak are now in good agreement with experiment.

The OH-stretch $\nu_1$ VFR is strongly suppressed by the availablity of 2-quantum vibrational levels, with $\Gamma _\nu^e/\Gamma_\nu =0.043$ (see Table~\ref{tab:VFR_ch3oh}). Analysis of its escape width $\Gamma _\nu^v$ shows that it decays predominantly into the $\nu_6+\nu_7$ final state (86\% of $\Gamma _\nu^v$) and $2\nu_6$ state (7\%). However, the suppression of this resonance is accompanied by the emergence of a 3-quantum combination VFR $2\nu _6+\nu_7$, with $\Gamma _\nu^e/\Gamma_\nu =0.656$. This appears to be a consequence of the Darling-Dennison resonance between $\nu_1$ and $2\nu _6+\nu_7$ (at 3865.22 and 3866.46~cm$^{-1}$ in the harmonic approximation), that is not accounted for by the present VPT2 approach.
We are thus dealing with a strongly mixed pair of levels. Of the two states, the $2\nu _6+\nu_7$ VFR has a larger elastic width, which suggests that the pure mode and combination labels should be swapped. In this case $\nu_1$ will be used for the VFR with the stronger coupling to the vibrational ground state, and $2\nu _6+\nu_7$ for the VFR with the large decay rates towards $\nu_6+\nu_7$ and $2\nu_6$ final states. Labeling aside, the final 1--3-quantum calculation provides a reasonable description of the measured OH-stretch peak, though is slightly smaller in magnitude.

The importance of overtones and combinations for the description of $\Z$ in methanol was proposed earlier in Ref.~\cite{YGLS08}. However, the conclusions drawn in that paper are correct only qualitatively. The measured $\Z$ spectrum presented in Ref.~\cite{YGLS08} suffered from errors and the theoretical treatment made use of rather uncertain absorption data obtained in liquid methanol  \cite{Bertie1997}.

As with other molecules, the calculated $\Z$ fails to describe the measured annihilation in the gaps between the VFRs of the modes and at higher energies. In methanol this unexplained signal is observed between 0.2 and 0.27 eV and above 0.5~eV. Theory also strongly underestimates the experimental data 
below 0.08~eV. We address this discrepancy below.

\subsection{Summary}

In summary, anharmonic corrections to the vibrational eigenstates and transition dipole amplitudes of all four molecules discussed in in Sec.~\ref{sec:res} have a pronounced effect on the calculated $\Z$.
The first and simplest change in comparison with the harmonic approximation is the shift of the vibrational state energies, which brings them into close agreement with experiment. This is particularly noticeable for the VFRs of the CH- and OH-stretch stretch modes in methanol.

Second, these anharmonic corrections enable direct coupling between the modes. As a result, a number of mode-based (i.e., single-quantum) VFR become suppressed due to vibrationally inelastic escape. This leads to an almost complete disappearance of the CH-stretch (or CD-stretch) resonances in the three chlorine-containing molecules. In these molecule, the CH-stretch modes have the smallest dipole coupling to the ground state, but couple more strongly to lower-lying modes, such as the CH bend or CH rocking. Previously, strong suppression of the CH-stretch peak (which is prominent in all alkanes with more than two carbon atoms) was observed experimentally in fluorine-substituted molecules \cite{Barnes2003,Barnes2006,YS08}.

Third, anharmonic effects allow positron capture in VFRs of overtones and combination vibrations. Significant contributions of 2-quantum VFRs are observed in all molecules, while 3-quantum resonances also contribute in chloroform, 1,1-dichloroethylene and methanol. The importance of multimode vibrational excitations was invoked in Ref.~\cite{Gribakin2000} in order to explain the strong enhancement of the annihilation rates in larger polyatomic molecules. The rapid increase of $\Z$ with the size of the molecule (e.g., for alkanes, see Refs.~\cite{Iwata1995} and \cite{Barnes2003} for room-temperature thermal and energy-resolved annihilation data, respectively) cannot be explained by considering only the VFRs of the fundamentals \cite{GG04,GL09,RMP2010}.
It is thus very important that the present calculations show how the multimode VFR are ``turned on'' by the anharmonic interactions.

It is interesting to note that most of the 2- and 3-quantum VFR that produce significant contributions to the $\Z$ spectrum (i.e., with $\Gamma _\nu^e/\Gamma_\nu \sim 1$) have energies close to one of the single-mode VFR (see Tables~\ref{tab:VFR_chcl3}, \ref{tab:VFR_cdcl3}, \ref{tab:VFR_c2h2cl2}, and \ref{tab:VFR_ch3oh}). As a result, they appear to enhance the magnitudes of mode-based resonances, rather than produce new features in the $\Z$ spectrum.
This behavior is similar to the observed enhancement of peaks in the measured $\Z$ spectra in larger polyatomic molecules (e.g., CH stretch in alkanes),
where increases in their heights are beyond that explicable by the VFRs of the modes \cite{Barnes2003,Barnes2006,YS08,GL09}. The effect is a consequence of perturbative mixing between the states, which is clearly stronger when their energies are close,
even in the absence of the profound mixings that accompany strong ``Fermi resonances'' and ``Darling-Dennison resonances''. In these cases, mode-based vibrational excitations serve as doorways into the dense spectrum of multimode VFRs \cite{GG04}. 

Figures \ref{fig:chcl3}, \ref{fig:cdcl3}, \ref{fig:c2h2cl2}, and \ref{fig:ch3oh} show that including the anharmonic effects brings the calculated resonant $\Z$ into closer agreement with experimental data for all four molecules, compared with the harmonic calculations which include only the VFRs of the modes. In all four cases the calculations reproduce the overall energy dependence of the measured annihilation rate. They also account for the magnitudes of the main peaks observed in the $\Z$ spectra, though with up to 30--50\% discrepancies in some cases.

\section{Multimode resonances}\label{sec:disc}

One feature that the above calculations fail to describe is the annihilation rate at energies between the VFR peaks and above the highest-frequency mode (e.g., the CH, CD, or OH stretch). This ``missing signal'' has the form of a smooth, slowly decreasing background that underlies the distinct VFR peaks. One mechanism that can produce such contribution is the direct, in-flight annihilation (see Ref.~\cite{Green2014} for a complete description of this phenomenon in noble gases). For atoms and molecules in which the positrons have a low-energy virtual state or a weakly bound state, the corresponding annihilation rate can be evaluated as \cite{Dzuba1993,Dzuba1996,Goldanskii1964}
\begin{equation}\label{eq:Zeffdir}
Z_\text{eff}^\text{(dir)}\simeq \frac{F}{\kappa ^2+k^2},
\end{equation}
where $\kappa =\sqrt{2\eps _b}$ and $F$ is the same constant as in Eq.~(\ref{eq:rho_ep}) \cite{Gribakin2001,RMP2010}. As an estimate, for 0.3~eV positrons and $\eps _b\lesssim 40$~meV, Eq.~(\ref{eq:Zeffdir}) gives $Z_\text{eff}^\text{(dir)}\approx 30$, which is smaller than the observed $\Z$ background.

Another mechanism that can produce such background in $\Z$ is the so-called multimode resonant annihilation (MRA). A statistical description of this phenomenon (SMRA) can be found in Refs.~\cite{GL09,Jones2012}. Its main idea is similar to resonant annihilation outlined in Sec.~\ref{subsec:rate}. However, SMRA considers the limit of dense vibrational spectra in which the levels are strongly mixed, and the contributions of individual resonances cannot be resolved. In this case, the SMRA contribution can be estimated as \cite{GL09}
\begin{equation}\label{eq:Zeffmra}
Z_\text{eff}^\text{(mra)}(\eps )=\frac{2\pi ^2\rho_{ep}}{k}\frac{\rho (\eps +E_{\nu_0}+\eps _b)}{N(\eps +E_{\nu_0})},
\end{equation}
where $\rho (E)$ is the density of the molecular vibrational states, $N(E)=\int_0^E\rho (E')dE'$ is the number of levels with energies up to $E$ ($E=0$ for the ground state), and $\nu _0$ is the initial vibrational state of the molecule. Application of Eq.~(\ref{eq:Zeffmra}) to alkanes, C$_n$H$_{2n+2}$, with 3 to 8 carbon atoms, showed that $Z_\text{eff}^\text{(mra)}$ does account for the annihilation rates observed between the mode-based peaks \cite{GL09}. A subsequent paper \cite{Jones2012} examined a number of smaller molecules (halomethanes CHCl$_3$, CCl$_4$, CHBr$_3$, and CBr$_4$). It found that the energy dependence of the measured annihilation rates could be explained by assuming a significant SMRA contribution, although $Z_\text{eff}^\text{(mra)}$ had to be scaled by a factor $\eta \lesssim 0.1$.

\begin{figure}
\includegraphics*[width=0.48\textwidth]{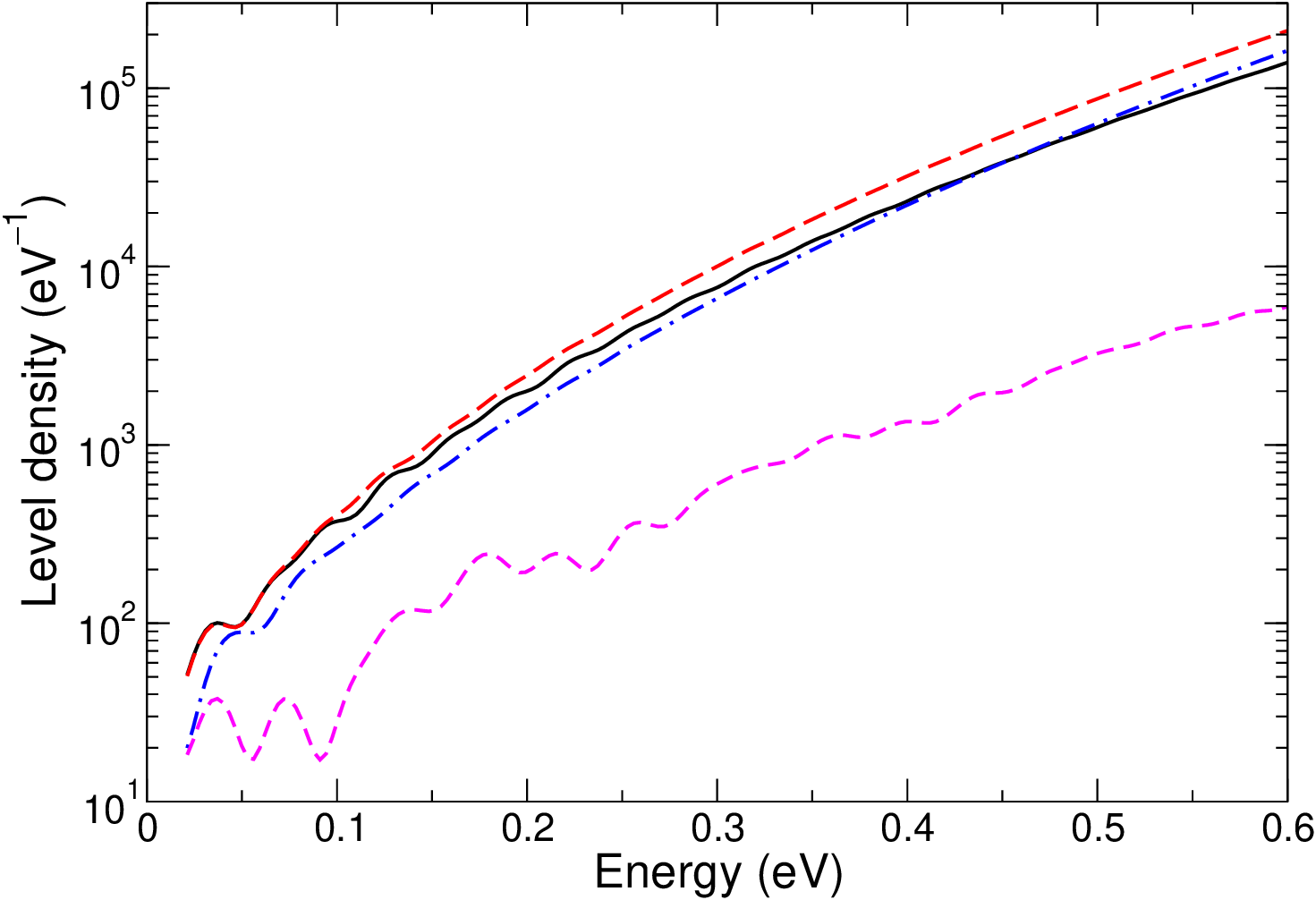}
\caption{Vibrational level densities $\rho (E)$ calculated in the harmonic approximation using mode frequencies from Ref.~\cite{NIST} for chloroform (solid line), chloroform-$d_1$ (long-dashed line), 1,1-dichloroethylene (dot-dashed line), and methanol (short-dashed line). For plotting, the densities have been folded with a Gaussian with 25~meV FWHM.}
\label{fig:lev_dens}
\end{figure}

To see whether the SMRA contribution can be significant in the molecules studied above, the experimental $\Z$ data is fit by the sum
\begin{equation}\label{eq:Zefftot}
\Z (\epsilon ) = \bar{Z}_{\rm eff}^{(\rm res)}(\epsilon )+\eta \bar Z_\text{eff}^\text{(mra)}(\epsilon ),
\end{equation}
where $\bar Z_\text{eff}^\text{(mra)}(\epsilon )$ is obtained by averaging Eq.~(\ref{eq:Zeffmra}) over the positron energy distribution,
\begin{equation}\label{eq:Zeffmra1}
\bar Z_\text{eff}^\text{(mra)}(\epsilon )=\int Z_\text{eff}^\text{(mra)}(\eps ') \Delta (\epsilon -\eps ')d\eps ',
\end{equation}
and $\eta $ is chosen to reproduce the signal away from the mode-based peaks. Figure~\ref{fig:lev_dens} shows the vibrational level densities used to calculate $Z_\text{eff}^\text{(mra)}(\eps )$. It was also averaged over the room-temperature Boltzmann distribution of the initial vibrational states $\nu_0$ of the molecule, although this had only a small effect on $\bar Z_\text{eff}^\text{(mra)}(\epsilon )$. For methanol, which has the highest vibrational frequencies and the smallest SMRA contribution, $Z_\text{eff}^\text{(dir)}$ was also added in Eq.~(\ref{eq:Zefftot}) when constructing the total $\Z$.

\begin{figure}
\includegraphics*[width=0.48\textwidth]{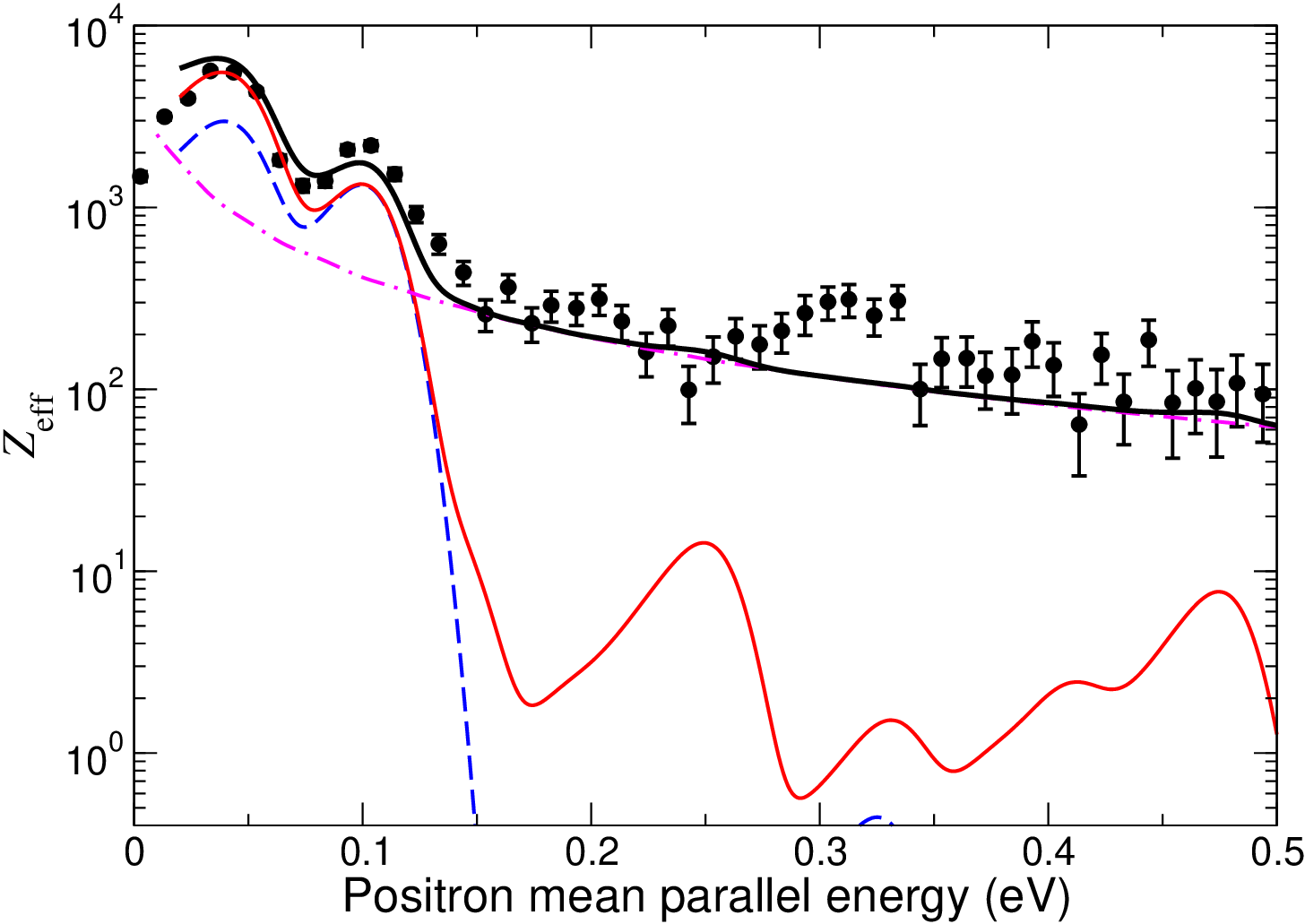}
\caption{Calculated and measured annihilation rate for chloroform:
dashed line, $\bar{Z}_{\rm eff}^{(\rm res)}$ due to mode-based VFR (anharmonic);
solid line, $\bar{Z}_{\rm eff}^{(\rm res)}$ due to 1--3-quantum VFR; dot-dashed line, 
SMRA $\Z$, Eq.~(\ref{eq:Zeffmra1}), scaled by $\eta =0.3$; thick solid line,
$\bar{Z}_{\rm eff}^{(\rm res)}+\eta \bar Z_\text{eff}^\text{(mra)}$; solid circles, measured $\Z$ \cite{Jones2013,Natisin_thesis}.}
\label{fig:chcl3_log}
\end{figure}

\begin{figure}
\includegraphics*[width=0.48\textwidth]{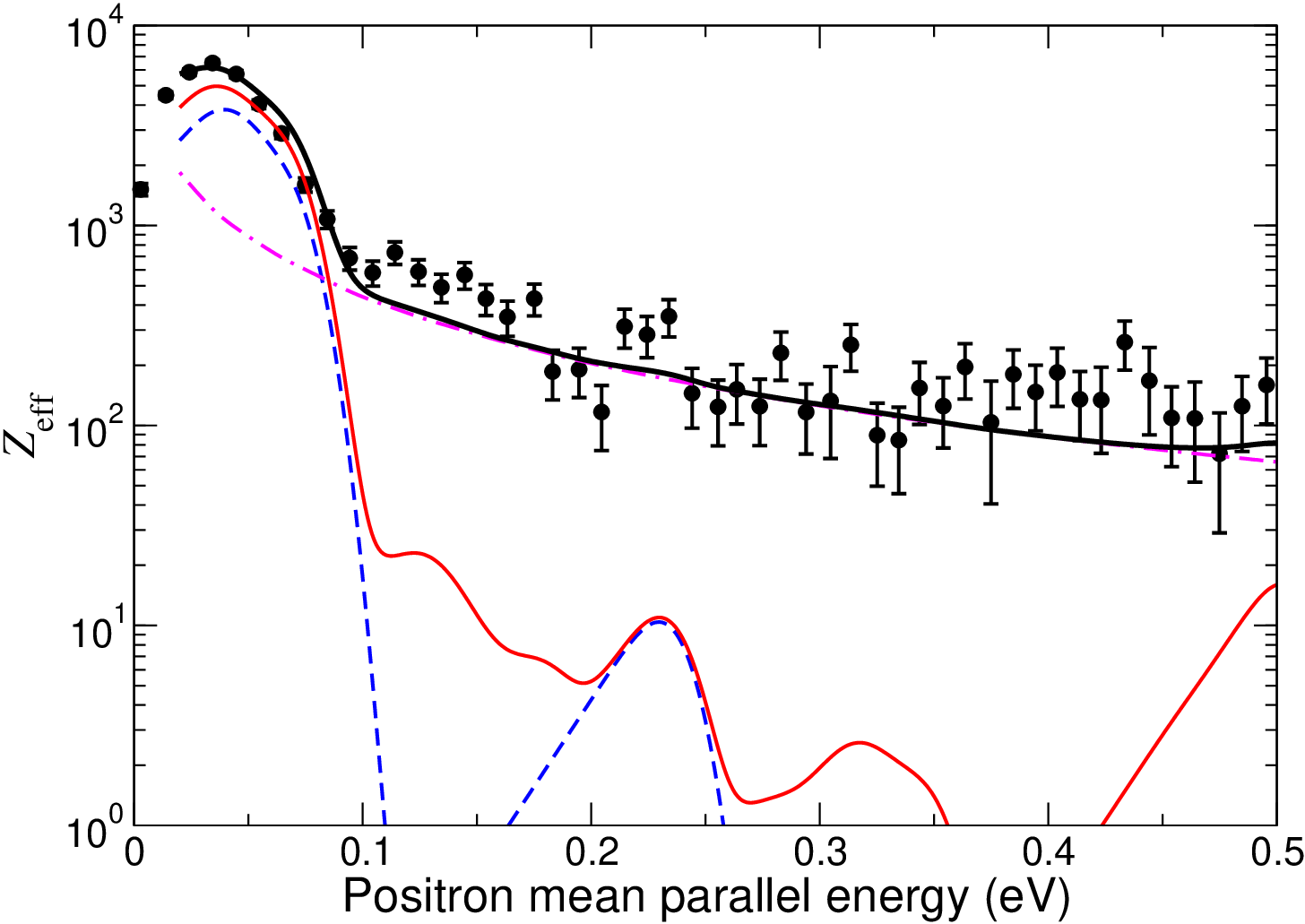}
\caption{Calculated and measured annihilation rate for chloroform-$d_1$:
dashed line, $\bar{Z}_{\rm eff}^{(\rm res)}$ due to mode-based VFR (anharmonic);
solid line, $\bar{Z}_{\rm eff}^{(\rm res)}$ due to 1--3-quantum VFR; dot-dashed line, 
SMRA $\Z$, Eq.~(\ref{eq:Zeffmra1}), scaled by $\eta =0.3$; thick solid line,
$\bar{Z}_{\rm eff}^{(\rm res)}+\eta \bar Z_\text{eff}^\text{(mra)}$; solid circles, measured $\Z$ \cite{Jones2013,Natisin_thesis}.}
\label{fig:cdcl3_log}
\end{figure}

\begin{figure}
\includegraphics*[width=0.48\textwidth]{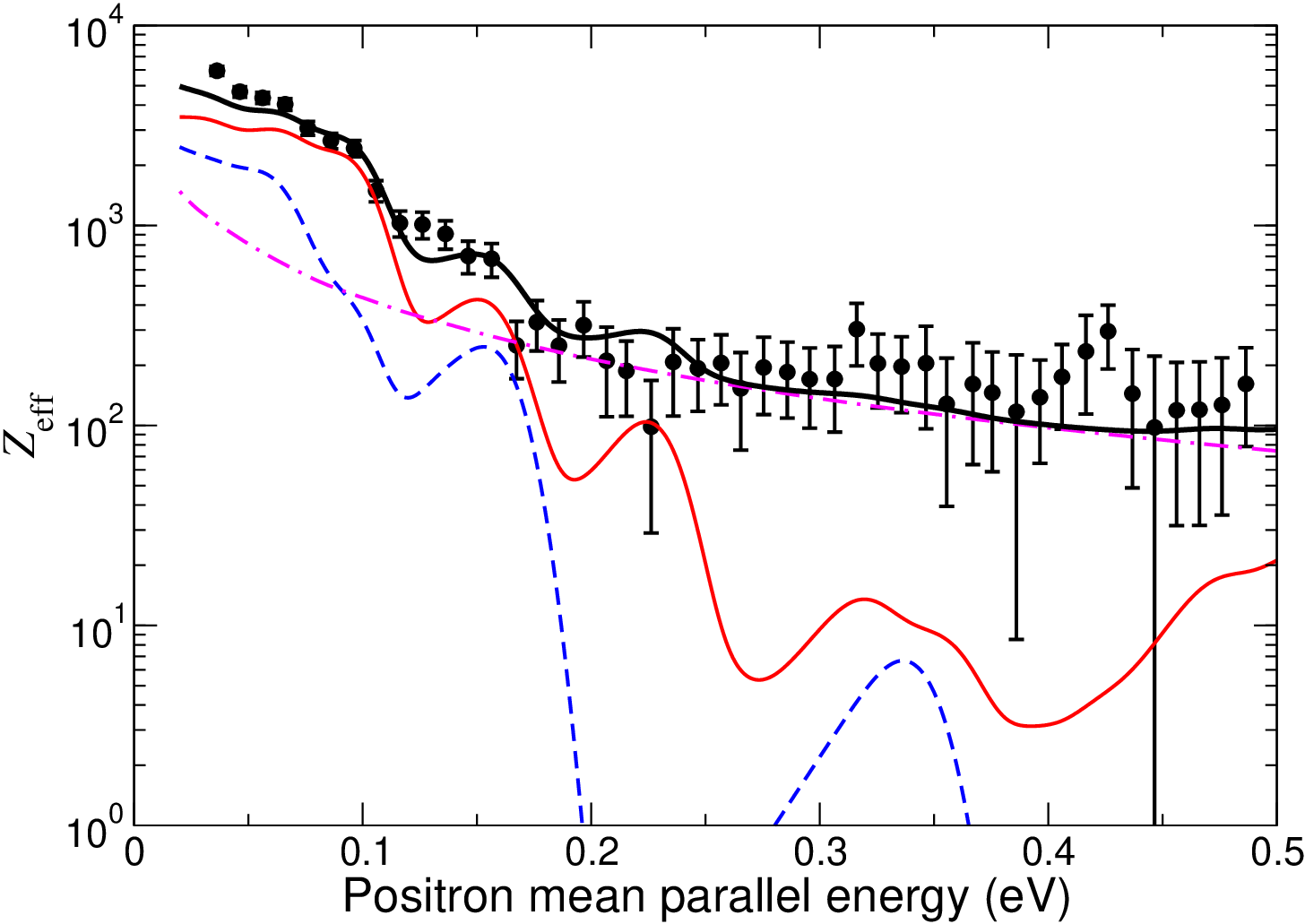}
\caption{Calculated and measured annihilation rate for 1,1-dichloroethylene:
dashed line, $\bar{Z}_{\rm eff}^{(\rm res)}$ due to mode-based VFR (anharmonic);
solid line, $\bar{Z}_{\rm eff}^{(\rm res)}$ due to 1--3-quantum VFR; dot-dashed line, 
SMRA $\Z$, Eq.~(\ref{eq:Zeffmra1}), scaled by $\eta =0.35$; thick solid line,
$\bar{Z}_{\rm eff}^{(\rm res)}+\eta \bar Z_\text{eff}^\text{(mra)}$; solid circles, measured $\Z$ \cite{Jones2013,Natisin_thesis}.}
\label{fig:c2h2cl2_log}
\end{figure}

\begin{figure}
\includegraphics*[width=0.48\textwidth]{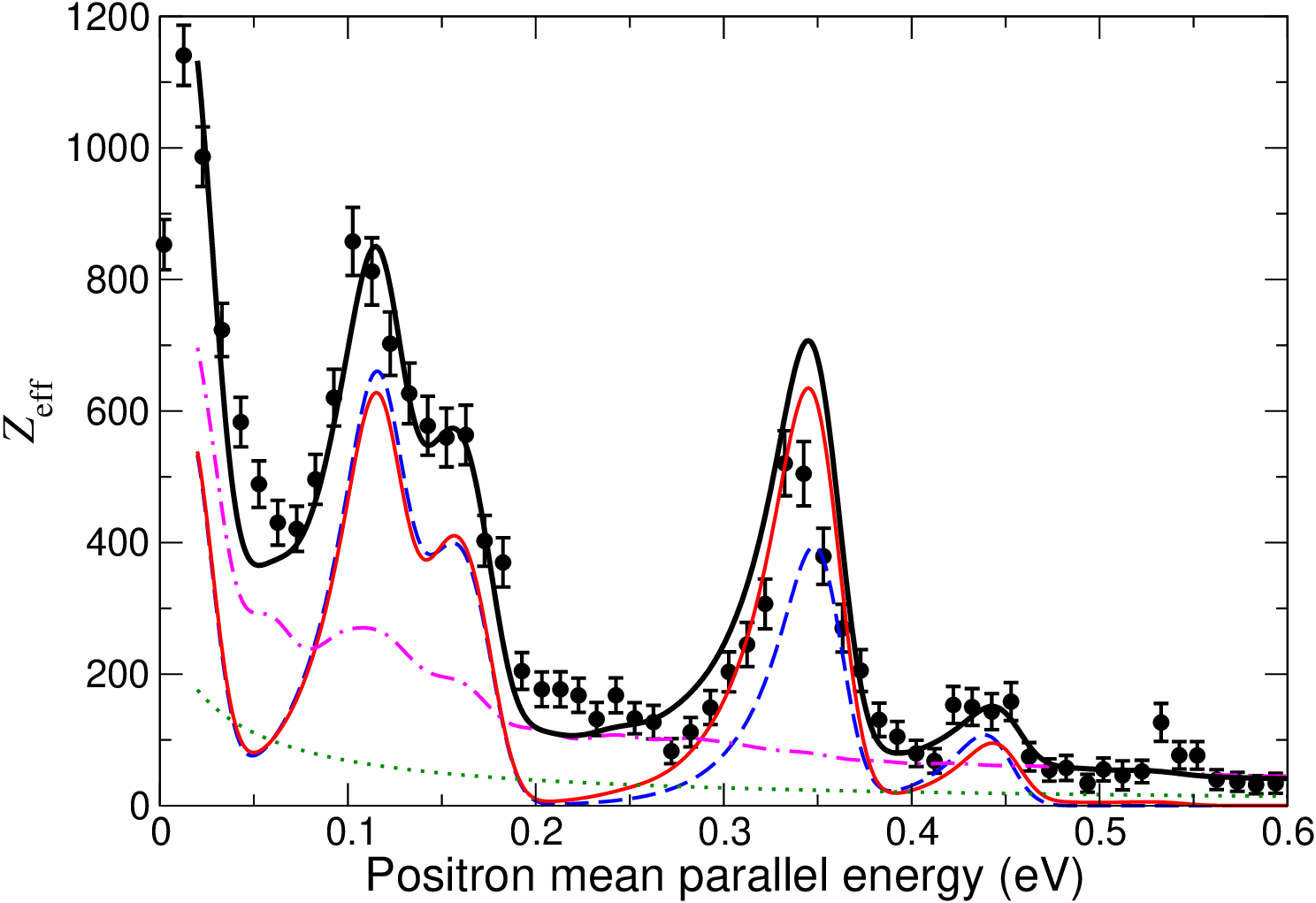}
\caption{Calculated and measured annihilation rate for methanol:
dashed line, $\bar{Z}_{\rm eff}^{(\rm res)}$ due to mode-based VFR (anharmonic);
solid line, $\bar{Z}_{\rm eff}^{(\rm res)}$ due to 1--3-quantum VFR; dot-dashed line, 
SMRA $\Z$, Eq.~(\ref{eq:Zeffmra1}); dotted line, direct $\Z$, Eq.~(\ref{eq:Zeffdir}); thick solid line, $\bar{Z}_{\rm eff}^{(\rm res)}+Z_{\rm eff}^{(\rm dir)}+\eta \bar Z_\text{eff}^\text{(mra)}$, with $\eta =0.6$; solid circles, measured $\Z$ \cite{Natisin_thesis}.}
\label{fig:ch3oh_1}
\end{figure}

As seen in Figs.~\ref{fig:chcl3_log}, \ref{fig:cdcl3_log}, \ref{fig:c2h2cl2_log}, and \ref{fig:ch3oh_1}, in all four molecules the SMRA contribution produced a distinct contribution to the measured annihilation signal. The fitted values of $\eta $ range from 0.3 in chloroform and chloroform-$d_1$, and 0.35 in 1,1-dichloroethylene to 0.6 in methanol. For the molecules containing chlorine, the $\Z$ are presented on a logarithmic scale. This enables one to see that 2- and 3-quantum VFRs do provide some contribution to the annihilation rate $\bar{Z}_{\rm eff}^{(\rm res)}$ at all energies. However, this contribution is insufficient to describe the measured $\Z$ above 0.1~eV in chloroform-$d_1$, 0.15~eV in chloroform and 0.2~eV in 1,1-dichloroethylene; at these energies the SMRA in fact dominates the signal.

Of the molecules considered in this work, methanol has the smallest $\Z$ values, because it has the smallest binding energy and lacks the low-energy vibrational modes that produce larger contributions to the signal. It also has by far the lowest vibrational level density (see Fig.~\ref{fig:lev_dens}), and the smallest SMRA contribution. Hence, in methanol we also include $\Z$ due to direct annihilation, Eq.~(\ref{eq:Zeffdir}), whose contribution is about 30\% of $\bar Z_\text{eff}^\text{(mra)}$. Figure~\ref{fig:ch3oh_1} shows that adding the SMRA and direct annihilation contributions to the resonant $\Z$ presented in Sec.~\ref{sec:res}, produces a near-perfect description of the measured $\Z$ spectrum, except for a small overestimation of the CH-stretch maximum.

As can be seen from Figs.~\ref{fig:chcl3_log}-\ref{fig:ch3oh_1}, adding the SMRA contribution results in an improved agreement between theory and experiment. However, the present treatment certainly lacks the rigor of the explicit \textit{ab initio} calculation of the annihilation rate due to 1--3-quantum VFRs, as described in Sec.~\ref{sec:theory} and presented in Sec.~\ref{sec:res}. Both $Z_{\rm eff}^{(\rm res)}$ and $Z_{\rm eff}^{(\rm mra)}$ describe the same physical phenomenon of resonant annihilation, and their separation is somewhat artificial; it is an acknowledgment of our current inability to account for the anharmonic effects in arbitrary multiquantum vibrational resonances. To perform a complete calculation for a 5- or 6-atomic molecule would be a formidable task, while doing this for much larger molecules is simply unfeasible, as well as being a daunting prospect from the theoretical point of view.

To justify the current, pragmatic approach, we have checked that the total vibrational spectral densities shown in Fig.~\ref{fig:lev_dens} are significantly greater than those that include only 1--3-quantum resonances, for energies larger than 0.15~eV in chloroform, chloroform-$d_1$, and 1,1-dichloroethylene, and larger than 0.2--0.3~eV in methanol. Hence, the possible double counting between $Z_{\rm eff}^{(\rm res)}$ and $Z_{\rm eff}^{(\rm mra)}$ is restricted to lower energies, where the former dominates for all three chlorine-containing molecules. In methanol, it is possible that the large contribution of $Z_{\rm eff}^{(\rm mra)}$ at low energies effectively makes up for the inaccurate handling of the strongly anharmonic low-energy torsion mode ($\nu_{12}$) by the calculations described in Secs.~\ref{subsec:vib}.

\section{Conclusions}

To conclude, we have shown that an accurate description of the vibrational spectrum and transition amplitudes that accounts for anharmonic effects enables one to calculate positron resonant annihilation rates for molecules of modest size (e.g.,  5 or 6 atoms). It produces results that are in good overall agreement with the experimental data, completely ameliorating the qualitative deficiencies observed when the simple harmonic oscillator treatment is applied.
The calculations demonstrate that anharmonic effects can suppress the magnitudes of some resonances due to the effect of vibrationally inelastic escape, while other peaks can be enhanced by the contributions of nearby 2- or 3-quantum vibrational resonances.
This work is a significant advance towards a complete theory of positron annihilation in polyatomic molecules. Below we outline some near-term developments that can be foreseen.

Besides increasing the size of the vibrational space and the order of anharmonic corrections included in the calculations, several other improvements in the theory are called for. A more complete theory should go beyond the long-range dipole coupling description of positron interactions with vibrations. Short-range interactions can have a noticeable effect on the transition amplitudes and corresponding widths, especially at larger positron energies and for the transitions with small dipole amplitudes. In fact, it has recently been shown that infrared-inactive vibrational excitations do produce distinct VFRs in the positron annihilation spectra \cite{Natisin2017}. Such calculations should also employ true positron bound-state and continuum wave functions, instead of the plane wave and approximate analytical wave functions
used in the present theory (Sec.~\ref{subsec:widths}).

On the experimental side, the development of the cryogenically cooled, high-energy-resolution, trap-based positron beam promises to produce much higher resolution $\Z$ spectra that can be expected to exhibit individual energy-resolved VFRs. These spectra will provide more stringent tests of the theory. In particular, it is possible that the high-resolution beam will enable at least some 2- or 3-quantum resonances to be observed directly and analyzed separately from the mode-based resonances.

\section*{Acknowledgments}

We wish to acknowledge insightful conversations with R. W. Field that led to this work. The work at UCSD was supported by the U. S. NSF, grants PHY 14-01794 and 17-02230. JFS thanks the US Department of Energy, Office of Basic Energy Sciences, under award DE-FG02-07ER15884.


\begin{thebibliography}{57}%
\makeatletter
\providecommand \@ifxundefined [1]{%
 \@ifx{#1\undefined}
}%
\providecommand \@ifnum [1]{%
 \ifnum #1\expandafter \@firstoftwo
 \else \expandafter \@secondoftwo
 \fi
}%
\providecommand \@ifx [1]{%
 \ifx #1\expandafter \@firstoftwo
 \else \expandafter \@secondoftwo
 \fi
}%
\providecommand \natexlab [1]{#1}%
\providecommand \enquote  [1]{``#1''}%
\providecommand \bibnamefont  [1]{#1}%
\providecommand \bibfnamefont [1]{#1}%
\providecommand \citenamefont [1]{#1}%
\providecommand \href@noop [0]{\@secondoftwo}%
\providecommand \href [0]{\begingroup \@sanitize@url \@href}%
\providecommand \@href[1]{\@@startlink{#1}\@@href}%
\providecommand \@@href[1]{\endgroup#1\@@endlink}%
\providecommand \@sanitize@url [0]{\catcode `\\12\catcode `\$12\catcode
  `\&12\catcode `\#12\catcode `\^12\catcode `\_12\catcode `\%12\relax}%
\providecommand \@@startlink[1]{}%
\providecommand \@@endlink[0]{}%
\providecommand \url  [0]{\begingroup\@sanitize@url \@url }%
\providecommand \@url [1]{\endgroup\@href {#1}{\urlprefix }}%
\providecommand \urlprefix  [0]{URL }%
\providecommand \Eprint [0]{\href }%
\providecommand \doibase [0]{http://dx.doi.org/}%
\providecommand \selectlanguage [0]{\@gobble}%
\providecommand \bibinfo  [0]{\@secondoftwo}%
\providecommand \bibfield  [0]{\@secondoftwo}%
\providecommand \translation [1]{[#1]}%
\providecommand \BibitemOpen [0]{}%
\providecommand \bibitemStop [0]{}%
\providecommand \bibitemNoStop [0]{.\EOS\space}%
\providecommand \EOS [0]{\spacefactor3000\relax}%
\providecommand \BibitemShut  [1]{\csname bibitem#1\endcsname}%
\let\auto@bib@innerbib\@empty
\bibitem [{\citenamefont {Gribakin}\ \emph {et~al.}(2010)\citenamefont
  {Gribakin}, \citenamefont {Young},\ and\ \citenamefont {Surko}}]{RMP2010}%
  \BibitemOpen
  \bibfield  {author} {\bibinfo {author} {\bibfnamefont {G.~F.}\ \bibnamefont
  {Gribakin}}, \bibinfo {author} {\bibfnamefont {J.~A.}\ \bibnamefont {Young}},
  \ and\ \bibinfo {author} {\bibfnamefont {C.~M.}\ \bibnamefont {Surko}},\
  }\href {\doibase 10.1103/RevModPhys.82.2557} {\bibfield  {journal} {\bibinfo
  {journal} {Rev. Mod. Phys.}\ }\textbf {\bibinfo {volume} {82}},\ \bibinfo
  {pages} {2557} (\bibinfo {year} {2010})}\BibitemShut {NoStop}%
\bibitem [{\citenamefont {Gribakin}(2000)}]{Gribakin2000}%
  \BibitemOpen
  \bibfield  {author} {\bibinfo {author} {\bibfnamefont {G.~F.}\ \bibnamefont
  {Gribakin}},\ }\href {\doibase 10.1103/PhysRevA.61.022720} {\bibfield
  {journal} {\bibinfo  {journal} {Phys. Rev. A}\ }\textbf {\bibinfo {volume}
  {61}},\ \bibinfo {pages} {022720} (\bibinfo {year} {2000})}\BibitemShut
  {NoStop}%
\bibitem [{\citenamefont {Gribakin}(2001)}]{Gribakin2001}%
  \BibitemOpen
  \bibfield  {author} {\bibinfo {author} {\bibfnamefont {G.~F.}\ \bibnamefont
  {Gribakin}},\ }in\ \href@noop {} {\emph {\bibinfo {booktitle} {New Directions
  in Antimatter Chemistry and Physics}}},\ \bibinfo {editor} {edited by\
  \bibinfo {editor} {\bibfnamefont {C.~M.}\ \bibnamefont {Surko}}\ and\
  \bibinfo {editor} {\bibfnamefont {F.~A.}\ \bibnamefont {Gianturco}}}\
  (\bibinfo  {publisher} {Kluwer Academic},\ \bibinfo {address} {Dordrecht},\
  \bibinfo {year} {2001})\ Chap.~\bibinfo {chapter} {22}\BibitemShut {NoStop}%
\bibitem [{\citenamefont {Gilbert}\ \emph {et~al.}(2002)\citenamefont
  {Gilbert}, \citenamefont {Barnes}, \citenamefont {Sullivan},\ and\
  \citenamefont {Surko}}]{Gilbert2002}%
  \BibitemOpen
  \bibfield  {author} {\bibinfo {author} {\bibfnamefont {S.~J.}\ \bibnamefont
  {Gilbert}}, \bibinfo {author} {\bibfnamefont {L.~D.}\ \bibnamefont {Barnes}},
  \bibinfo {author} {\bibfnamefont {J.~P.}\ \bibnamefont {Sullivan}}, \ and\
  \bibinfo {author} {\bibfnamefont {C.~M.}\ \bibnamefont {Surko}},\ }\href
  {\doibase 10.1103/PhysRevLett.88.043201} {\bibfield  {journal} {\bibinfo
  {journal} {Phys. Rev. Lett.}\ }\textbf {\bibinfo {volume} {88}},\ \bibinfo
  {pages} {043201} (\bibinfo {year} {2002})}\BibitemShut {NoStop}%
\bibitem [{\citenamefont {Barnes}\ \emph {et~al.}(2003)\citenamefont {Barnes},
  \citenamefont {Gilbert},\ and\ \citenamefont {Surko}}]{Barnes2003}%
  \BibitemOpen
  \bibfield  {author} {\bibinfo {author} {\bibfnamefont {L.~D.}\ \bibnamefont
  {Barnes}}, \bibinfo {author} {\bibfnamefont {S.~J.}\ \bibnamefont {Gilbert}},
  \ and\ \bibinfo {author} {\bibfnamefont {C.~M.}\ \bibnamefont {Surko}},\
  }\href {\doibase 10.1103/PhysRevA.67.032706} {\bibfield  {journal} {\bibinfo
  {journal} {Phys. Rev. A}\ }\textbf {\bibinfo {volume} {67}},\ \bibinfo
  {pages} {032706} (\bibinfo {year} {2003})}\BibitemShut {NoStop}%
\bibitem [{\citenamefont {Danielson}\ \emph {et~al.}(2009)\citenamefont
  {Danielson}, \citenamefont {Young},\ and\ \citenamefont
  {Surko}}]{Danielson2009}%
  \BibitemOpen
  \bibfield  {author} {\bibinfo {author} {\bibfnamefont {J.~R.}\ \bibnamefont
  {Danielson}}, \bibinfo {author} {\bibfnamefont {J.~A.}\ \bibnamefont
  {Young}}, \ and\ \bibinfo {author} {\bibfnamefont {C.~M.}\ \bibnamefont
  {Surko}},\ }\href {\doibase 10.1088/0953-4075/42/23/235203} {\bibfield
  {journal} {\bibinfo  {journal} {J. Phys. B}\ }\textbf {\bibinfo {volume}
  {42}},\ \bibinfo {pages} {235203} (\bibinfo {year} {2009})}\BibitemShut
  {NoStop}%
\bibitem [{\citenamefont {Danielson}\ \emph {et~al.}(2010)\citenamefont
  {Danielson}, \citenamefont {Gosselin},\ and\ \citenamefont
  {Surko}}]{Danielson2010}%
  \BibitemOpen
  \bibfield  {author} {\bibinfo {author} {\bibfnamefont {J.~R.}\ \bibnamefont
  {Danielson}}, \bibinfo {author} {\bibfnamefont {J.~J.}\ \bibnamefont
  {Gosselin}}, \ and\ \bibinfo {author} {\bibfnamefont {C.~M.}\ \bibnamefont
  {Surko}},\ }\href {\doibase 10.1103/PhysRevLett.104.233201} {\bibfield
  {journal} {\bibinfo  {journal} {Phys. Rev. Lett.}\ }\textbf {\bibinfo
  {volume} {104}},\ \bibinfo {pages} {233201} (\bibinfo {year}
  {2010})}\BibitemShut {NoStop}%
\bibitem [{\citenamefont {Danielson}\ \emph
  {et~al.}(2012{\natexlab{a}})\citenamefont {Danielson}, \citenamefont {Jones},
  \citenamefont {Gosselin}, \citenamefont {Natisin},\ and\ \citenamefont
  {Surko}}]{Danielson2012a}%
  \BibitemOpen
  \bibfield  {author} {\bibinfo {author} {\bibfnamefont {J.~R.}\ \bibnamefont
  {Danielson}}, \bibinfo {author} {\bibfnamefont {A.~C.~L.}\ \bibnamefont
  {Jones}}, \bibinfo {author} {\bibfnamefont {J.~J.}\ \bibnamefont {Gosselin}},
  \bibinfo {author} {\bibfnamefont {M.~R.}\ \bibnamefont {Natisin}}, \ and\
  \bibinfo {author} {\bibfnamefont {C.~M.}\ \bibnamefont {Surko}},\ }\href
  {\doibase 10.1103/PhysRevA.85.022709} {\bibfield  {journal} {\bibinfo
  {journal} {Phys. Rev. A}\ }\textbf {\bibinfo {volume} {85}},\ \bibinfo
  {pages} {022709} (\bibinfo {year} {2012}{\natexlab{a}})}\BibitemShut
  {NoStop}%
\bibitem [{\citenamefont {Danielson}\ \emph
  {et~al.}(2012{\natexlab{b}})\citenamefont {Danielson}, \citenamefont {Jones},
  \citenamefont {Natisin},\ and\ \citenamefont {Surko}}]{Danielson2012b}%
  \BibitemOpen
  \bibfield  {author} {\bibinfo {author} {\bibfnamefont {J.~R.}\ \bibnamefont
  {Danielson}}, \bibinfo {author} {\bibfnamefont {A.~C.~L.}\ \bibnamefont
  {Jones}}, \bibinfo {author} {\bibfnamefont {M.~R.}\ \bibnamefont {Natisin}},
  \ and\ \bibinfo {author} {\bibfnamefont {C.~M.}\ \bibnamefont {Surko}},\
  }\href {\doibase 10.1103/PhysRevLett.109.113201} {\bibfield  {journal}
  {\bibinfo  {journal} {Phys. Rev. Lett.}\ }\textbf {\bibinfo {volume} {109}},\
  \bibinfo {pages} {113201} (\bibinfo {year} {2012}{\natexlab{b}})}\BibitemShut
  {NoStop}%
\bibitem [{\citenamefont {Gribakin}\ and\ \citenamefont
  {Lee}(2006)}]{Gribakin2006}%
  \BibitemOpen
  \bibfield  {author} {\bibinfo {author} {\bibfnamefont {G.~F.}\ \bibnamefont
  {Gribakin}}\ and\ \bibinfo {author} {\bibfnamefont {C.~M.~R.}\ \bibnamefont
  {Lee}},\ }\href {\doibase 10.1103/PhysRevLett.97.193201} {\bibfield
  {journal} {\bibinfo  {journal} {Phys. Rev. Lett.}\ }\textbf {\bibinfo
  {volume} {97}},\ \bibinfo {pages} {193201} (\bibinfo {year}
  {2006})}\BibitemShut {NoStop}%
\bibitem [{\citenamefont {Tachikawa}\ \emph {et~al.}(2011)\citenamefont
  {Tachikawa}, \citenamefont {Kita},\ and\ \citenamefont {Buenker}}]{TKB11}%
  \BibitemOpen
  \bibfield  {author} {\bibinfo {author} {\bibfnamefont {M.}~\bibnamefont
  {Tachikawa}}, \bibinfo {author} {\bibfnamefont {Y.}~\bibnamefont {Kita}}, \
  and\ \bibinfo {author} {\bibfnamefont {R.~J.}\ \bibnamefont {Buenker}},\
  }\href {\doibase 10.1039/C0CP01650K} {\bibfield  {journal} {\bibinfo
  {journal} {Phys. Chem. Chem. Phys.}\ }\textbf {\bibinfo {volume} {13}},\
  \bibinfo {pages} {2701} (\bibinfo {year} {2011})}\BibitemShut {NoStop}%
\bibitem [{\citenamefont {Tachikawa}\ \emph {et~al.}(2012)\citenamefont
  {Tachikawa}, \citenamefont {Kita},\ and\ \citenamefont {Buenker}}]{TKB12}%
  \BibitemOpen
  \bibfield  {author} {\bibinfo {author} {\bibfnamefont {M.}~\bibnamefont
  {Tachikawa}}, \bibinfo {author} {\bibfnamefont {Y.}~\bibnamefont {Kita}}, \
  and\ \bibinfo {author} {\bibfnamefont {R.~J.}\ \bibnamefont {Buenker}},\
  }\href {\doibase 10.1088/1367-2630/14/3/035004} {\bibfield  {journal}
  {\bibinfo  {journal} {New J. Phys.}\ }\textbf {\bibinfo {volume} {14}},\
  \bibinfo {pages} {035004} (\bibinfo {year} {2012})}\BibitemShut {NoStop}%
\bibitem [{\citenamefont {Tachikawa}(2014)}]{T14}%
  \BibitemOpen
  \bibfield  {author} {\bibinfo {author} {\bibfnamefont {M.}~\bibnamefont
  {Tachikawa}},\ }\href {\doibase 10.1088/1742-6596/488/1/012053} {\bibfield
  {journal} {\bibinfo  {journal} {J. Phys. Conf. Ser.}\ }\textbf {\bibinfo
  {volume} {488}},\ \bibinfo {pages} {012053} (\bibinfo {year}
  {2014})}\BibitemShut {NoStop}%
\bibitem [{\citenamefont {Young}\ and\ \citenamefont
  {Surko}(2008{\natexlab{a}})}]{YS08}%
  \BibitemOpen
  \bibfield  {author} {\bibinfo {author} {\bibfnamefont {J.~A.}\ \bibnamefont
  {Young}}\ and\ \bibinfo {author} {\bibfnamefont {C.~M.}\ \bibnamefont
  {Surko}},\ }\href {\doibase 10.1103/PhysRevA.77.052704} {\bibfield  {journal}
  {\bibinfo  {journal} {Phys. Rev. A}\ }\textbf {\bibinfo {volume} {77}},\
  \bibinfo {pages} {052704} (\bibinfo {year} {2008}{\natexlab{a}})}\BibitemShut
  {NoStop}%
\bibitem [{\citenamefont {Gribakin}\ and\ \citenamefont {Lee}(2009)}]{GL09}%
  \BibitemOpen
  \bibfield  {author} {\bibinfo {author} {\bibfnamefont {G.~F.}\ \bibnamefont
  {Gribakin}}\ and\ \bibinfo {author} {\bibfnamefont {C.~M.~R.}\ \bibnamefont
  {Lee}},\ }\href {\doibase 10.1140/epjd/e2008-00188-9} {\bibfield  {journal}
  {\bibinfo  {journal} {Eur. Phys. J. D}\ }\textbf {\bibinfo {volume} {51}},\
  \bibinfo {pages} {51} (\bibinfo {year} {2009})}\BibitemShut {NoStop}%
\bibitem [{\citenamefont {Gribakin}\ and\ \citenamefont {Gill}(2004)}]{GG04}%
  \BibitemOpen
  \bibfield  {author} {\bibinfo {author} {\bibfnamefont {G.~F.}\ \bibnamefont
  {Gribakin}}\ and\ \bibinfo {author} {\bibfnamefont {P.~M.~W.}\ \bibnamefont
  {Gill}},\ }\href {\doibase 10.1016/j.nimb.2004.03.027} {\bibfield  {journal}
  {\bibinfo  {journal} {Nucl. Instrum. Methods Phys. Res., Sect. B}\ }\textbf
  {\bibinfo {volume} {221}},\ \bibinfo {pages} {30} (\bibinfo {year}
  {2004})}\BibitemShut {NoStop}%
\bibitem [{\citenamefont {Young}\ \emph {et~al.}(2008)\citenamefont {Young},
  \citenamefont {Gribakin}, \citenamefont {Lee},\ and\ \citenamefont
  {Surko}}]{YGLS08}%
  \BibitemOpen
  \bibfield  {author} {\bibinfo {author} {\bibfnamefont {J.~A.}\ \bibnamefont
  {Young}}, \bibinfo {author} {\bibfnamefont {G.~F.}\ \bibnamefont {Gribakin}},
  \bibinfo {author} {\bibfnamefont {C.~M.~R.}\ \bibnamefont {Lee}}, \ and\
  \bibinfo {author} {\bibfnamefont {C.~M.}\ \bibnamefont {Surko}},\ }\href
  {\doibase 10.1103/PhysRevA.77.060702} {\bibfield  {journal} {\bibinfo
  {journal} {Phys. Rev. A}\ }\textbf {\bibinfo {volume} {77}},\ \bibinfo
  {pages} {060702} (\bibinfo {year} {2008})}\BibitemShut {NoStop}%
\bibitem [{\citenamefont {Young}\ and\ \citenamefont
  {Surko}(2008{\natexlab{b}})}]{YS08a}%
  \BibitemOpen
  \bibfield  {author} {\bibinfo {author} {\bibfnamefont {J.~A.}\ \bibnamefont
  {Young}}\ and\ \bibinfo {author} {\bibfnamefont {C.~M.}\ \bibnamefont
  {Surko}},\ }\href {\doibase 10.1103/PhysRevA.78.032702} {\bibfield  {journal}
  {\bibinfo  {journal} {Phys. Rev. A}\ }\textbf {\bibinfo {volume} {78}},\
  \bibinfo {pages} {032702} (\bibinfo {year} {2008}{\natexlab{b}})}\BibitemShut
  {NoStop}%
\bibitem [{\citenamefont {Gribakin}(2010)}]{Gribakin2010}%
  \BibitemOpen
  \bibfield  {author} {\bibinfo {author} {\bibfnamefont {G.~F.}\ \bibnamefont
  {Gribakin}},\ }\href {\doibase 10.1088/1742-6596/199/1/012013} {\bibfield
  {journal} {\bibinfo  {journal} {J. Phys.: Conf. Ser.}\ }\textbf {\bibinfo
  {volume} {199}},\ \bibinfo {pages} {012013} (\bibinfo {year}
  {2010})}\BibitemShut {NoStop}%
\bibitem [{\citenamefont {Natisin}\ \emph {et~al.}(2017)\citenamefont
  {Natisin}, \citenamefont {Danielson}, \citenamefont {Gribakin}, \citenamefont
  {Swann},\ and\ \citenamefont {Surko}}]{Natisin2017}%
  \BibitemOpen
  \bibfield  {author} {\bibinfo {author} {\bibfnamefont {M.~R.}\ \bibnamefont
  {Natisin}}, \bibinfo {author} {\bibfnamefont {J.~R.}\ \bibnamefont
  {Danielson}}, \bibinfo {author} {\bibfnamefont {G.~F.}\ \bibnamefont
  {Gribakin}}, \bibinfo {author} {\bibfnamefont {A.~R.}\ \bibnamefont {Swann}},
  \ and\ \bibinfo {author} {\bibfnamefont {C.~M.}\ \bibnamefont {Surko}},\
  }\href {\doibase 10.1103/PhysRevLett.119.113402} {\bibfield  {journal}
  {\bibinfo  {journal} {Phys. Rev. Lett.}\ }\textbf {\bibinfo {volume} {119}},\
  \bibinfo {pages} {113402} (\bibinfo {year} {2017})}\BibitemShut {NoStop}%
\bibitem [{\citenamefont {Jones}\ \emph {et~al.}(2013)\citenamefont {Jones},
  \citenamefont {Danielson}, \citenamefont {Natisin},\ and\ \citenamefont
  {Surko}}]{Jones2013}%
  \BibitemOpen
  \bibfield  {author} {\bibinfo {author} {\bibfnamefont {A.~C.~L.}\
  \bibnamefont {Jones}}, \bibinfo {author} {\bibfnamefont {J.~R.}\ \bibnamefont
  {Danielson}}, \bibinfo {author} {\bibfnamefont {M.~R.}\ \bibnamefont
  {Natisin}}, \ and\ \bibinfo {author} {\bibfnamefont {C.~M.}\ \bibnamefont
  {Surko}},\ }\href {\doibase 10.1103/PhysRevLett.110.223201} {\bibfield
  {journal} {\bibinfo  {journal} {Phys. Rev. Lett.}\ }\textbf {\bibinfo
  {volume} {110}},\ \bibinfo {pages} {223201} (\bibinfo {year}
  {2013})}\BibitemShut {NoStop}%
\bibitem [{\citenamefont {Danielson}\ \emph {et~al.}(2013)\citenamefont
  {Danielson}, \citenamefont {Jones}, \citenamefont {Natisin},\ and\
  \citenamefont {Surko}}]{Danielson2013}%
  \BibitemOpen
  \bibfield  {author} {\bibinfo {author} {\bibfnamefont {J.~R.}\ \bibnamefont
  {Danielson}}, \bibinfo {author} {\bibfnamefont {A.~C.~L.}\ \bibnamefont
  {Jones}}, \bibinfo {author} {\bibfnamefont {M.~R.}\ \bibnamefont {Natisin}},
  \ and\ \bibinfo {author} {\bibfnamefont {C.~M.}\ \bibnamefont {Surko}},\
  }\href {\doibase 10.1103/PhysRevA.88.062702} {\bibfield  {journal} {\bibinfo
  {journal} {Phys. Rev. A}\ }\textbf {\bibinfo {volume} {88}},\ \bibinfo
  {pages} {062702} (\bibinfo {year} {2013})}\BibitemShut {NoStop}%
\bibitem [{\citenamefont {{Kita, Yukiumi}}\ and\ \citenamefont {{Tachikawa,
  Masanori}}(2014)}]{Kita2014}%
  \BibitemOpen
  \bibfield  {author} {\bibinfo {author} {\bibnamefont {{Kita, Yukiumi}}}\ and\
  \bibinfo {author} {\bibnamefont {{Tachikawa, Masanori}}},\ }\href {\doibase
  10.1140/epjd/e2014-40799-9} {\bibfield  {journal} {\bibinfo  {journal} {Eur.
  Phys. J. D}\ }\textbf {\bibinfo {volume} {68}},\ \bibinfo {pages} {116}
  (\bibinfo {year} {2014})}\BibitemShut {NoStop}%
\bibitem [{\citenamefont {Landau}\ and\ \citenamefont
  {Lifshitz}(1977)}]{LandauQM}%
  \BibitemOpen
  \bibfield  {author} {\bibinfo {author} {\bibfnamefont {L.~D.}\ \bibnamefont
  {Landau}}\ and\ \bibinfo {author} {\bibfnamefont {E.~M.}\ \bibnamefont
  {Lifshitz}},\ }\href@noop {} {\emph {\bibinfo {title} {Quantum Mechanics:
  Non-Relativistic Theory}}},\ \bibinfo {edition} {3rd}\ ed.\ (\bibinfo
  {publisher} {Pergamon},\ \bibinfo {address} {Oxford},\ \bibinfo {year}
  {1977})\BibitemShut {NoStop}%
\bibitem [{\citenamefont {Gilbert}\ \emph {et~al.}(1997)\citenamefont
  {Gilbert}, \citenamefont {Kurz}, \citenamefont {Greaves},\ and\ \citenamefont
  {Surko}}]{Gilbert1997}%
  \BibitemOpen
  \bibfield  {author} {\bibinfo {author} {\bibfnamefont {S.~J.}\ \bibnamefont
  {Gilbert}}, \bibinfo {author} {\bibfnamefont {C.}~\bibnamefont {Kurz}},
  \bibinfo {author} {\bibfnamefont {R.~G.}\ \bibnamefont {Greaves}}, \ and\
  \bibinfo {author} {\bibfnamefont {C.~M.}\ \bibnamefont {Surko}},\ }\href
  {\doibase 10.1063/1.118787} {\bibfield  {journal} {\bibinfo  {journal}
  {Applied Physics Letters}\ }\textbf {\bibinfo {volume} {70}},\ \bibinfo
  {pages} {1944} (\bibinfo {year} {1997})}\BibitemShut {NoStop}%
\bibitem [{\citenamefont {Natisin}\ \emph {et~al.}(2016)\citenamefont
  {Natisin}, \citenamefont {Danielson},\ and\ \citenamefont
  {Surko}}]{Natisin2016}%
  \BibitemOpen
  \bibfield  {author} {\bibinfo {author} {\bibfnamefont {M.~R.}\ \bibnamefont
  {Natisin}}, \bibinfo {author} {\bibfnamefont {J.~R.}\ \bibnamefont
  {Danielson}}, \ and\ \bibinfo {author} {\bibfnamefont {C.~M.}\ \bibnamefont
  {Surko}},\ }\href {\doibase 10.1063/1.4939854} {\bibfield  {journal}
  {\bibinfo  {journal} {Appl. Phys. Lett.}\ }\textbf {\bibinfo {volume}
  {108}},\ \bibinfo {pages} {024102} (\bibinfo {year} {2016})}\BibitemShut
  {NoStop}%
\bibitem [{Note1()}]{Note1}%
  \BibitemOpen
  \bibinfo {note} {The hypergeometric function in Eq.~(\ref {eq:h}) is
  $_2F_1\left (\protect \frac {1}{2},1;\protect \frac {5}{2};-z^2\right )
  =\protect \frac 32 z^{-2}[(1+z^2)z^{-1}\protect \qopname \relax o{arctan}z
  -1]$.}\BibitemShut {Stop}%
\bibitem [{\citenamefont {Mitroy}\ and\ \citenamefont
  {Ivanov}(2002)}]{Mitroy2002}%
  \BibitemOpen
  \bibfield  {author} {\bibinfo {author} {\bibfnamefont {J.}~\bibnamefont
  {Mitroy}}\ and\ \bibinfo {author} {\bibfnamefont {I.~A.}\ \bibnamefont
  {Ivanov}},\ }\href {\doibase 10.1103/PhysRevA.65.042705} {\bibfield
  {journal} {\bibinfo  {journal} {Phys. Rev. A}\ }\textbf {\bibinfo {volume}
  {65}},\ \bibinfo {pages} {042705} (\bibinfo {year} {2002})}\BibitemShut
  {NoStop}%
\bibitem [{\citenamefont {Fraser}(1968)}]{Fraser68}%
  \BibitemOpen
  \bibfield  {author} {\bibinfo {author} {\bibfnamefont {P.~A.}\ \bibnamefont
  {Fraser}},\ }\href {\doibase 10.1016/S0065-2199(08)60185-2} {\bibfield
  {journal} {\bibinfo  {journal} {Adv. At. Mol. Phys.}\ }\textbf {\bibinfo
  {volume} {4}},\ \bibinfo {pages} {63} (\bibinfo {year} {1968})}\BibitemShut
  {NoStop}%
\bibitem [{\citenamefont {Deutsch}(1951)}]{Deutsch51}%
  \BibitemOpen
  \bibfield  {author} {\bibinfo {author} {\bibfnamefont {M.}~\bibnamefont
  {Deutsch}},\ }\href {\doibase 10.1103/PhysRev.83.866} {\bibfield  {journal}
  {\bibinfo  {journal} {Phys. Rev.}\ }\textbf {\bibinfo {volume} {83}},\
  \bibinfo {pages} {866} (\bibinfo {year} {1951})}\BibitemShut {NoStop}%
\bibitem [{Note2()}]{Note2}%
  \BibitemOpen
  \bibinfo {note} {Note that a factor $1/2$ is missing in the expression for
  $\Delta (E)$ in Ref.~\cite {Gribakin2006}, Eq.~(10), although is was included
  in the calculations.}\BibitemShut {Stop}%
\bibitem [{\citenamefont {Mills}(1972)}]{VPT2}%
  \BibitemOpen
  \bibfield  {author} {\bibinfo {author} {\bibfnamefont {I.~M.}\ \bibnamefont
  {Mills}},\ }in\ \href@noop {} {\emph {\bibinfo {booktitle} {Modern
  Spectroscopy: Modern Research}}},\ \bibinfo {editor} {edited by\ \bibinfo
  {editor} {\bibfnamefont {K.~N.}\ \bibnamefont {Rao}}\ and\ \bibinfo {editor}
  {\bibfnamefont {C.~W.}\ \bibnamefont {Matthews}}}\ (\bibinfo  {publisher}
  {Academic, New York},\ \bibinfo {year} {1972})\ p.\ \bibinfo {pages}
  {115}\BibitemShut {NoStop}%
\bibitem [{\citenamefont {Raghavachari}\ \emph {et~al.}(1989)\citenamefont
  {Raghavachari}, \citenamefont {Trucks}, \citenamefont {Pople},\ and\
  \citenamefont {Head-Gordon}}]{CC}%
  \BibitemOpen
  \bibfield  {author} {\bibinfo {author} {\bibfnamefont {K.}~\bibnamefont
  {Raghavachari}}, \bibinfo {author} {\bibfnamefont {G.~W.}\ \bibnamefont
  {Trucks}}, \bibinfo {author} {\bibfnamefont {J.~A.}\ \bibnamefont {Pople}}, \
  and\ \bibinfo {author} {\bibfnamefont {M.}~\bibnamefont {Head-Gordon}},\
  }\href {\doibase 10.1016/S0009-2614(89)87395-6} {\bibfield  {journal}
  {\bibinfo  {journal} {Chem. Phys. Lett.}\ }\textbf {\bibinfo {volume}
  {157}},\ \bibinfo {pages} {479 } (\bibinfo {year} {1989})}\BibitemShut
  {NoStop}%
\bibitem [{\citenamefont {Feierabend}\ \emph {et~al.}(2006)\citenamefont
  {Feierabend}, \citenamefont {Havey}, \citenamefont {Varner}, \citenamefont
  {Stanton},\ and\ \citenamefont {Vaida}}]{THREEA}%
  \BibitemOpen
  \bibfield  {author} {\bibinfo {author} {\bibfnamefont {K.~J.}\ \bibnamefont
  {Feierabend}}, \bibinfo {author} {\bibfnamefont {D.~K.}\ \bibnamefont
  {Havey}}, \bibinfo {author} {\bibfnamefont {M.~E.}\ \bibnamefont {Varner}},
  \bibinfo {author} {\bibfnamefont {J.~F.}\ \bibnamefont {Stanton}}, \ and\
  \bibinfo {author} {\bibfnamefont {V.}~\bibnamefont {Vaida}},\ }\href
  {\doibase 10.1063/1.2180248} {\bibfield  {journal} {\bibinfo  {journal} {J.
  Chem. Phys.}\ }\textbf {\bibinfo {volume} {124}},\ \bibinfo {pages} {124323}
  (\bibinfo {year} {2006})}\BibitemShut {NoStop}%
\bibitem [{\citenamefont {Bloino}(2015)}]{THREEB}%
  \BibitemOpen
  \bibfield  {author} {\bibinfo {author} {\bibfnamefont {J.}~\bibnamefont
  {Bloino}},\ }\href {\doibase 10.1021/jp509985u} {\bibfield  {journal}
  {\bibinfo  {journal} {J. Phys. Chem. A}\ }\textbf {\bibinfo {volume} {119}},\
  \bibinfo {pages} {5269} (\bibinfo {year} {2015})}\BibitemShut {NoStop}%
\bibitem [{\citenamefont {Jones}\ \emph
  {et~al.}(2012{\natexlab{a}})\citenamefont {Jones}, \citenamefont {Danielson},
  \citenamefont {Natisin}, \citenamefont {Surko},\ and\ \citenamefont
  {Gribakin}}]{Jones2012}%
  \BibitemOpen
  \bibfield  {author} {\bibinfo {author} {\bibfnamefont {A.~C.~L.}\
  \bibnamefont {Jones}}, \bibinfo {author} {\bibfnamefont {J.~R.}\ \bibnamefont
  {Danielson}}, \bibinfo {author} {\bibfnamefont {M.~R.}\ \bibnamefont
  {Natisin}}, \bibinfo {author} {\bibfnamefont {C.~M.}\ \bibnamefont {Surko}},
  \ and\ \bibinfo {author} {\bibfnamefont {G.~F.}\ \bibnamefont {Gribakin}},\
  }\href {\doibase 10.1103/PhysRevLett.108.093201} {\bibfield  {journal}
  {\bibinfo  {journal} {Phys. Rev. Lett.}\ }\textbf {\bibinfo {volume} {108}},\
  \bibinfo {pages} {093201} (\bibinfo {year} {2012}{\natexlab{a}})}\BibitemShut
  {NoStop}%
\bibitem [{\citenamefont {McCaslin}\ and\ \citenamefont
  {Stanton}(2013)}]{ANOA}%
  \BibitemOpen
  \bibfield  {author} {\bibinfo {author} {\bibfnamefont {L.}~\bibnamefont
  {McCaslin}}\ and\ \bibinfo {author} {\bibfnamefont {J.}~\bibnamefont
  {Stanton}},\ }\href {\doibase 10.1080/00268976.2013.811303} {\bibfield
  {journal} {\bibinfo  {journal} {Mol. Phys.}\ }\textbf {\bibinfo {volume}
  {111}},\ \bibinfo {pages} {1492} (\bibinfo {year} {2013})}\BibitemShut
  {NoStop}%
\bibitem [{\citenamefont {Alml\"of}\ and\ \citenamefont {Taylor}(1987)}]{ANOB}%
  \BibitemOpen
  \bibfield  {author} {\bibinfo {author} {\bibfnamefont {J.}~\bibnamefont
  {Alml\"of}}\ and\ \bibinfo {author} {\bibfnamefont {P.~R.}\ \bibnamefont
  {Taylor}},\ }\href {\doibase 10.1063/1.451917} {\bibfield  {journal}
  {\bibinfo  {journal} {J. Chem. Phys.}\ }\textbf {\bibinfo {volume} {86}},\
  \bibinfo {pages} {4070} (\bibinfo {year} {1987})}\BibitemShut {NoStop}%
\bibitem [{\citenamefont {Dzuba}\ \emph {et~al.}(1995)\citenamefont {Dzuba},
  \citenamefont {Flambaum}, \citenamefont {Gribakin},\ and\ \citenamefont
  {King}}]{Dzuba1995}%
  \BibitemOpen
  \bibfield  {author} {\bibinfo {author} {\bibfnamefont {V.~A.}\ \bibnamefont
  {Dzuba}}, \bibinfo {author} {\bibfnamefont {V.~V.}\ \bibnamefont {Flambaum}},
  \bibinfo {author} {\bibfnamefont {G.~F.}\ \bibnamefont {Gribakin}}, \ and\
  \bibinfo {author} {\bibfnamefont {W.~A.}\ \bibnamefont {King}},\ }\href
  {\doibase 10.1103/PhysRevA.52.4541} {\bibfield  {journal} {\bibinfo
  {journal} {Phys. Rev. A}\ }\textbf {\bibinfo {volume} {52}},\ \bibinfo
  {pages} {4541} (\bibinfo {year} {1995})}\BibitemShut {NoStop}%
\bibitem [{\citenamefont {Mitroy}\ \emph {et~al.}(1999)\citenamefont {Mitroy},
  \citenamefont {Bromley},\ and\ \citenamefont {Ryzhikh}}]{Mitroy1999}%
  \BibitemOpen
  \bibfield  {author} {\bibinfo {author} {\bibfnamefont {J.}~\bibnamefont
  {Mitroy}}, \bibinfo {author} {\bibfnamefont {M.~W.~J.}\ \bibnamefont
  {Bromley}}, \ and\ \bibinfo {author} {\bibfnamefont {G.}~\bibnamefont
  {Ryzhikh}},\ }\href {\doibase 10.1088/0953-4075/32/9/311} {\bibfield
  {journal} {\bibinfo  {journal} {J. Phys. B}\ }\textbf {\bibinfo {volume}
  {32}},\ \bibinfo {pages} {2203} (\bibinfo {year} {1999})}\BibitemShut
  {NoStop}%
\bibitem [{\citenamefont {Dzuba}\ \emph {et~al.}(2010)\citenamefont {Dzuba},
  \citenamefont {Flambaum},\ and\ \citenamefont {Gribakin}}]{Dzuba2010}%
  \BibitemOpen
  \bibfield  {author} {\bibinfo {author} {\bibfnamefont {V.~A.}\ \bibnamefont
  {Dzuba}}, \bibinfo {author} {\bibfnamefont {V.~V.}\ \bibnamefont {Flambaum}},
  \ and\ \bibinfo {author} {\bibfnamefont {G.~F.}\ \bibnamefont {Gribakin}},\
  }\href {\doibase 10.1103/PhysRevLett.105.203401} {\bibfield  {journal}
  {\bibinfo  {journal} {Phys. Rev. Lett.}\ }\textbf {\bibinfo {volume} {105}},\
  \bibinfo {pages} {203401} (\bibinfo {year} {2010})}\BibitemShut {NoStop}%
\bibitem [{\citenamefont {Gribakin}\ and\ \citenamefont
  {Swann}(2015)}]{Gribakin2015}%
  \BibitemOpen
  \bibfield  {author} {\bibinfo {author} {\bibfnamefont {G.~F.}\ \bibnamefont
  {Gribakin}}\ and\ \bibinfo {author} {\bibfnamefont {A.~R.}\ \bibnamefont
  {Swann}},\ }\href {\doibase 10.1088/0953-4075/48/21/215101} {\bibfield
  {journal} {\bibinfo  {journal} {J. Phys. B}\ }\textbf {\bibinfo {volume}
  {48}},\ \bibinfo {pages} {215101} (\bibinfo {year} {2015})}\BibitemShut
  {NoStop}%
\bibitem [{\citenamefont {Natisin}(2016)}]{Natisin_thesis}%
  \BibitemOpen
  \bibfield  {author} {\bibinfo {author} {\bibfnamefont {M.~R.}\ \bibnamefont
  {Natisin}},\ }\href@noop {} {Ph.D. thesis},\ \bibinfo  {school} {University
  of California, San Diego} (\bibinfo {year} {2016})\BibitemShut {NoStop}%
\bibitem [{\citenamefont {Lide}(2005)}]{CRCHandbook}%
  \BibitemOpen
  \bibinfo {editor} {\bibfnamefont {D.~R.}\ \bibnamefont {Lide}},\ ed.,\
  \href@noop {} {\emph {\bibinfo {title} {CRC Handbook of Chemistry and
  Physics}}}\ (\bibinfo  {publisher} {CRC Press},\ \bibinfo {address} {Boca
  Raton, FL},\ \bibinfo {year} {2005})\BibitemShut {NoStop}%
\bibitem [{\citenamefont {Linstrom}\ and\ \citenamefont
  {Mallard}(2017)}]{NIST}%
  \BibitemOpen
  \bibinfo {editor} {\bibfnamefont {P.}~\bibnamefont {Linstrom}}\ and\ \bibinfo
  {editor} {\bibfnamefont {W.}~\bibnamefont {Mallard}},\ eds.,\ \href {\doibase
  10.18434/T4D303} {\emph {\bibinfo {title} {NIST Chemistry WebBook, NIST
  Standard Reference Database Number 69}}}\ (\bibinfo  {publisher} {NIST},\
  \bibinfo {address} {Gaithersburg MD},\ \bibinfo {year} {2017})\BibitemShut
  {NoStop}%
\bibitem [{\citenamefont {Bishop}\ and\ \citenamefont
  {Cheung}(1982)}]{Bishop1982}%
  \BibitemOpen
  \bibfield  {author} {\bibinfo {author} {\bibfnamefont {D.~M.}\ \bibnamefont
  {Bishop}}\ and\ \bibinfo {author} {\bibfnamefont {L.~M.}\ \bibnamefont
  {Cheung}},\ }\href {\doibase 10.1063/1.555658} {\bibfield  {journal}
  {\bibinfo  {journal} {J. Phys. Chem. Ref. Data}\ }\textbf {\bibinfo {volume}
  {11}},\ \bibinfo {pages} {119} (\bibinfo {year} {1982})}\BibitemShut
  {NoStop}%
\bibitem [{\citenamefont {Nishida}\ \emph {et~al.}(2012)\citenamefont
  {Nishida}, \citenamefont {Shigeto}, \citenamefont {Yabumoto},\ and\
  \citenamefont {o~Hamaguchi}}]{Nishida2012}%
  \BibitemOpen
  \bibfield  {author} {\bibinfo {author} {\bibfnamefont {J.}~\bibnamefont
  {Nishida}}, \bibinfo {author} {\bibfnamefont {S.}~\bibnamefont {Shigeto}},
  \bibinfo {author} {\bibfnamefont {S.}~\bibnamefont {Yabumoto}}, \ and\
  \bibinfo {author} {\bibfnamefont {H.}~\bibnamefont {o~Hamaguchi}},\ }\href
  {\doibase 10.1063/1.4770264} {\bibfield  {journal} {\bibinfo  {journal} {J.
  Chem. Phys.}\ }\textbf {\bibinfo {volume} {137}},\ \bibinfo {pages} {234501}
  (\bibinfo {year} {2012})}\BibitemShut {NoStop}%
\bibitem [{\citenamefont {Jones}\ \emph
  {et~al.}(2012{\natexlab{b}})\citenamefont {Jones}, \citenamefont {Danielson},
  \citenamefont {Gosselin}, \citenamefont {Natisin},\ and\ \citenamefont
  {Surko}}]{Jones2012a}%
  \BibitemOpen
  \bibfield  {author} {\bibinfo {author} {\bibfnamefont {A.~C.~L.}\
  \bibnamefont {Jones}}, \bibinfo {author} {\bibfnamefont {J.~R.}\ \bibnamefont
  {Danielson}}, \bibinfo {author} {\bibfnamefont {J.~J.}\ \bibnamefont
  {Gosselin}}, \bibinfo {author} {\bibfnamefont {M.~R.}\ \bibnamefont
  {Natisin}}, \ and\ \bibinfo {author} {\bibfnamefont {C.~M.}\ \bibnamefont
  {Surko}},\ }\href {\doibase 10.1088/1367-2630/14/1/015006} {\bibfield
  {journal} {\bibinfo  {journal} {New J. Phys.}\ }\textbf {\bibinfo {volume}
  {14}},\ \bibinfo {pages} {015006} (\bibinfo {year}
  {2012}{\natexlab{b}})}\BibitemShut {NoStop}%
\bibitem [{\citenamefont {Bertie}\ and\ \citenamefont
  {Zhang}(1997)}]{Bertie1997}%
  \BibitemOpen
  \bibfield  {author} {\bibinfo {author} {\bibfnamefont {J.~E.}\ \bibnamefont
  {Bertie}}\ and\ \bibinfo {author} {\bibfnamefont {S.~L.}\ \bibnamefont
  {Zhang}},\ }\href {\doibase 10.1016/S0022-2860(97)00152-X} {\bibfield
  {journal} {\bibinfo  {journal} {J. Mol. Struct.}\ }\textbf {\bibinfo {volume}
  {413}},\ \bibinfo {pages} {333 } (\bibinfo {year} {1997})}\BibitemShut
  {NoStop}%
\bibitem [{\citenamefont {Florian}\ \emph {et~al.}(1997)\citenamefont
  {Florian}, \citenamefont {Leszczynski}, \citenamefont {Johnson},\ and\
  \citenamefont {Goodman}}]{Florian1997}%
  \BibitemOpen
  \bibfield  {author} {\bibinfo {author} {\bibfnamefont {J.}~\bibnamefont
  {Florian}}, \bibinfo {author} {\bibfnamefont {J.}~\bibnamefont
  {Leszczynski}}, \bibinfo {author} {\bibfnamefont {B.~G.}\ \bibnamefont
  {Johnson}}, \ and\ \bibinfo {author} {\bibfnamefont {L.}~\bibnamefont
  {Goodman}},\ }\href {\doibase 10.1080/002689797171337} {\bibfield  {journal}
  {\bibinfo  {journal} {Mol. Phys.}\ }\textbf {\bibinfo {volume} {91}},\
  \bibinfo {pages} {439} (\bibinfo {year} {1997})}\BibitemShut {NoStop}%
\bibitem [{\citenamefont {Dang-Nhu}\ \emph {et~al.}(1990)\citenamefont
  {Dang-Nhu}, \citenamefont {Blanquet}, \citenamefont {Walrand}, \citenamefont
  {Allegrini},\ and\ \citenamefont {Moruzzi}}]{Dangnhu1990}%
  \BibitemOpen
  \bibfield  {author} {\bibinfo {author} {\bibfnamefont {M.}~\bibnamefont
  {Dang-Nhu}}, \bibinfo {author} {\bibfnamefont {G.}~\bibnamefont {Blanquet}},
  \bibinfo {author} {\bibfnamefont {J.}~\bibnamefont {Walrand}}, \bibinfo
  {author} {\bibfnamefont {M.}~\bibnamefont {Allegrini}}, \ and\ \bibinfo
  {author} {\bibfnamefont {G.}~\bibnamefont {Moruzzi}},\ }\href {\doibase
  10.1016/0022-2852(90)90172-M} {\bibfield  {journal} {\bibinfo  {journal} {J.
  Mol. Spectrosc.}\ }\textbf {\bibinfo {volume} {141}},\ \bibinfo {pages} {348
  } (\bibinfo {year} {1990})}\BibitemShut {NoStop}%
\bibitem [{\citenamefont {Barnes}\ \emph {et~al.}(2006)\citenamefont {Barnes},
  \citenamefont {Young},\ and\ \citenamefont {Surko}}]{Barnes2006}%
  \BibitemOpen
  \bibfield  {author} {\bibinfo {author} {\bibfnamefont {L.~D.}\ \bibnamefont
  {Barnes}}, \bibinfo {author} {\bibfnamefont {J.~A.}\ \bibnamefont {Young}}, \
  and\ \bibinfo {author} {\bibfnamefont {C.~M.}\ \bibnamefont {Surko}},\ }\href
  {\doibase 10.1103/PhysRevA.74.012706} {\bibfield  {journal} {\bibinfo
  {journal} {Phys. Rev. A}\ }\textbf {\bibinfo {volume} {74}},\ \bibinfo
  {pages} {012706} (\bibinfo {year} {2006})}\BibitemShut {NoStop}%
\bibitem [{\citenamefont {Iwata}\ \emph {et~al.}(1995)\citenamefont {Iwata},
  \citenamefont {Greaves}, \citenamefont {Murphy}, \citenamefont {Tinkle},\
  and\ \citenamefont {Surko}}]{Iwata1995}%
  \BibitemOpen
  \bibfield  {author} {\bibinfo {author} {\bibfnamefont {K.}~\bibnamefont
  {Iwata}}, \bibinfo {author} {\bibfnamefont {R.~G.}\ \bibnamefont {Greaves}},
  \bibinfo {author} {\bibfnamefont {T.~J.}\ \bibnamefont {Murphy}}, \bibinfo
  {author} {\bibfnamefont {M.~D.}\ \bibnamefont {Tinkle}}, \ and\ \bibinfo
  {author} {\bibfnamefont {C.~M.}\ \bibnamefont {Surko}},\ }\href {\doibase
  10.1103/PhysRevA.51.473} {\bibfield  {journal} {\bibinfo  {journal} {Phys.
  Rev. A}\ }\textbf {\bibinfo {volume} {51}},\ \bibinfo {pages} {473} (\bibinfo
  {year} {1995})}\BibitemShut {NoStop}%
\bibitem [{\citenamefont {Green}\ \emph {et~al.}(2014)\citenamefont {Green},
  \citenamefont {Ludlow},\ and\ \citenamefont {Gribakin}}]{Green2014}%
  \BibitemOpen
  \bibfield  {author} {\bibinfo {author} {\bibfnamefont {D.~G.}\ \bibnamefont
  {Green}}, \bibinfo {author} {\bibfnamefont {J.~A.}\ \bibnamefont {Ludlow}}, \
  and\ \bibinfo {author} {\bibfnamefont {G.~F.}\ \bibnamefont {Gribakin}},\
  }\href {\doibase 10.1103/PhysRevA.90.032712} {\bibfield  {journal} {\bibinfo
  {journal} {Phys. Rev. A}\ }\textbf {\bibinfo {volume} {90}},\ \bibinfo
  {pages} {032712} (\bibinfo {year} {2014})}\BibitemShut {NoStop}%
\bibitem [{\citenamefont {Dzuba}\ \emph {et~al.}(1993)\citenamefont {Dzuba},
  \citenamefont {Flambaum}, \citenamefont {King}, \citenamefont {Miller},\ and\
  \citenamefont {Sushkov}}]{Dzuba1993}%
  \BibitemOpen
  \bibfield  {author} {\bibinfo {author} {\bibfnamefont {V.~A.}\ \bibnamefont
  {Dzuba}}, \bibinfo {author} {\bibfnamefont {V.~V.}\ \bibnamefont {Flambaum}},
  \bibinfo {author} {\bibfnamefont {W.~A.}\ \bibnamefont {King}}, \bibinfo
  {author} {\bibfnamefont {B.~N.}\ \bibnamefont {Miller}}, \ and\ \bibinfo
  {author} {\bibfnamefont {O.~P.}\ \bibnamefont {Sushkov}},\ }\href {\doibase
  10.1088/0031-8949/1993/T46/039} {\bibfield  {journal} {\bibinfo  {journal}
  {Phys, Scripta}\ }\textbf {\bibinfo {volume} {1993}},\ \bibinfo {pages} {248}
  (\bibinfo {year} {1993})}\BibitemShut {NoStop}%
\bibitem [{\citenamefont {Dzuba}\ \emph {et~al.}(1996)\citenamefont {Dzuba},
  \citenamefont {Flambaum}, \citenamefont {Gribakin},\ and\ \citenamefont
  {King}}]{Dzuba1996}%
  \BibitemOpen
  \bibfield  {author} {\bibinfo {author} {\bibfnamefont {V.~A.}\ \bibnamefont
  {Dzuba}}, \bibinfo {author} {\bibfnamefont {V.~V.}\ \bibnamefont {Flambaum}},
  \bibinfo {author} {\bibfnamefont {G.~F.}\ \bibnamefont {Gribakin}}, \ and\
  \bibinfo {author} {\bibfnamefont {W.~A.}\ \bibnamefont {King}},\ }\href
  {\doibase 10.1088/0953-4075/29/14/024} {\bibfield  {journal} {\bibinfo
  {journal} {J. Phys. B}\ }\textbf {\bibinfo {volume} {29}},\ \bibinfo {pages}
  {3151} (\bibinfo {year} {1996})}\BibitemShut {NoStop}%
\bibitem [{\citenamefont {Goldanskii}\ and\ \citenamefont
  {Sayasov}(1964)}]{Goldanskii1964}%
  \BibitemOpen
  \bibfield  {author} {\bibinfo {author} {\bibfnamefont {V.}~\bibnamefont
  {Goldanskii}}\ and\ \bibinfo {author} {\bibfnamefont {Y.}~\bibnamefont
  {Sayasov}},\ }\href {\doibase http://dx.doi.org/10.1016/0031-9163(64)90018-6}
  {\bibfield  {journal} {\bibinfo  {journal} {Phys. Lett.}\ }\textbf {\bibinfo
  {volume} {13}},\ \bibinfo {pages} {300 } (\bibinfo {year}
  {1964})}\BibitemShut {NoStop}%
\end{thebibliography}

%

\end{document}